\documentclass[twocolumn]{aastex63}
\usepackage{CJKutf8} 
\usepackage{amsmath}

\shortauthors{Zaritsky, et al.}
\shorttitle{The Nature of UDGs}

\begin{document}

\title{Systematically Measuring Ultra-Diffuse Galaxies (SMUDGes). V.  The Complete SMUDGes Catalog and the Nature of Ultra-Diffuse Galaxies}
  
\correspondingauthor{Dennis Zaritsky}
\email{dennis.zaritsky@gmail.com}

\author[0000-0002-5177-727X]{Dennis Zaritsky}
\affiliation{Steward Observatory and Department of Astronomy, University of Arizona, 933 N. Cherry Ave., Tucson, AZ 85721, USA}

\author[0000-0001-7618-8212]{Richard Donnerstein}
\affiliation{Steward Observatory and Department of Astronomy, University of Arizona, 933 N. Cherry Ave., Tucson, AZ 85721, USA}

\author[0000-0002-4928-4003]{Arjun Dey}
\affiliation{NOIRLab, 950 N. Cherry Ave., Tucson, AZ 85719, USA}

\author[0000-0001-8855-3635]{Ananthan Karunakaran}
\affiliation{Instituto de Astrof\'{i}sica de Andaluc\'{i}a (CSIC), Glorieta de la Astronom\'{i}a, 18008 Granada, Spain}

\author[0000-0002-3767-9681]{Jennifer Kadowaki}
\affiliation{Steward Observatory and Department of Astronomy, University of Arizona, 933 N. Cherry Ave., Tucson, AZ 85721, USA}

\author[0000-0002-7013-4392]{Donghyeon J. Khim}
\affiliation{Steward Observatory and Department of Astronomy, University of Arizona, 933 N. Cherry Ave., Tucson, AZ 85721, USA}

\author[0000-0002-0956-7949]{Kristine Spekkens}
\affiliation{Department of Physics, Engineering Physics and Astronomy Queen's University Kingston, ON K7L 3N6, Canada}
\affiliation{Department of Physics and Space Science Royal Military College of Canada P.O. Box 17000, Station Forces Kingston, ON K7K 7B4, Canada}

\author[0000-0002-0123-9246]{Huanian Zhang\begin{CJK*}{UTF8}{gkai}(张华年)\end{CJK*}}
\affiliation{Department of Astronomy, Huazhong University of Science and Technology, Wuhan, Hubei 430074, China;}
\affiliation{Steward Observatory and Department of Astronomy, University of Arizona, 933 N. Cherry Ave., Tucson, AZ 85721, USA}

\begin{abstract}
We present the completed catalog of ultra-diffuse galaxy (UDG) candidates (7070 objects) from our search of the DR9 Legacy Survey images, including distance and total mass estimates for 1529 and 1436 galaxies, respectively, that we provide and describe in detail.
From the sample with estimated distances, we obtain a sample of 585 UDGs ($\mu_{0,g} \ge 24$ mag arcsec$^{-2}$ and $r_e \ge 1.5$ kpc) over 20,000 sq.\ deg of sky in various environments. We conclude that UDGs in our sample are limited to $10^{10} \lesssim$ M$_h$/M$_\odot  \lesssim 10^{11.5}$ and are on average a factor of 1.5 to 7 deficient in stars relative to the general population of galaxies of the same total mass. That factor increases with increasing galaxy size and mass up to a factor of $\sim$10 when the total mass of the UDG increases beyond M$_h = 10^{11}$ M$_\odot$. We do not find evidence that this factor has a dependence on the UDGs large-scale environment.
\end{abstract}


\section{Introduction}
\label{sec:intro}

The flurry of activity, both observational and theoretical in nature, since the \cite{vanDokkum+2015} study that coined the term ultra-diffuse galaxies (UDGs) to describe physically large low surface brightness galaxies,
has focused on understanding how these systems form and whether those processes highlight novel galaxy formation pathways or reflect extreme forms of already known phenomena.
Challenges in resolving this question include: 1) a  lack of homogeneous UDG data across environments; 
2) the possibly heterogeneous nature of the objects selected using  arbitrary physical size and surface brightness criteria \citep[see][for attempts to define more physically-motivated criteria]{trujillo20,li22}; and 3) a paucity of constraints on the underlying dark matter halos that host these systems. 

The first of these challenges we address with our search for UDGs in the images provided by the Dark Energy Spectroscopic Instrument (DESI) Legacy Imaging Surveys \citep[hereafter referred  to  as  the Legacy Survey;]
[]{Dey+2019}. We have described the basic principles of our work with the survey data and provided candidate UDG catalogs across subsets of the data in three papers \citep[][hereafter Papers I, II, and III]{Zaritsky+2019,Zaritsky+2021,Zaritsky+2022}. We refer to the survey as SMUDGes, which stems from the survey's full title, Systematically Measuring Ultra-Diffuse Galaxies. Here we present our complete catalog by augmenting a previous release of our analysis of the southern portion of the Legacy Survey (Paper III) with an analysis of the northern portion that we describe here in detail. Following most of the previous literature, we primarily identify candidates by applying a criterion based on central surface brightness in the $g$-band, $\mu_0 \ge 24$ mag arcsec$^{-2}$. Specific to this survey, we also require candidates to have an angular half-light radius, $r_e \ge 5.3$  arcsec. This peculiar size limit was set to correspond to a physical $r_e = 2.5\,$kpc at the distance of the Coma cluster, the environment explored by the first UDG surveys \citep{vanDokkum+2015,Koda+2015,Yagi+2016}. Unfortunately, selecting on angular size but defining UDGs in terms of physical units exacerbates the problem that any sample of UDG candidates is a heterogeneous population of galaxies.

The heterogeneity of the UDG samples and the initial overestimated masses of some UDGs \citep[cf.,][]{vanDokkum+2016} has led to apparently conflicting results regarding the classification of UDGs. Studies of individual UDGs \citep[e.g.,][]{Beasley+2016a,toloba,vdk19,forbes21}, which naturally focus on the largest, brightest objects, tend to conclude that these are relatively massive galaxies \citep[with total masses within the virial radius, including baryonic and dark matter, M$_h$, roughly comparable to, or larger than,  that of the Large Magellanic Cloud, M$_h \sim 1.4\times 10^{11}$ M$_\odot$;][]{erkal}, while studies using statistical samples of UDGs \citep[e.g.,][]{Beasley+2016b,amorisco}, which naturally focus on the more numerous smaller objects, tend to conclude that UDGs are lower-mass  galaxies (M$_h < 10^{11}$ M$_\odot$). However, the discrepancy is sometimes just a matter of emphasis. For example, the result of \cite{sifon}, who placed a statistical limit on UDG halo masses from gravitational lensing of log(M$_{200}/$M$_{\odot}) \le 11.8$ with high confidence, is sometimes cited as falling in the low mass camp simply because it is compared to the initial claims that some UDGs could have M$_h\ge$ 10$^{12}$ M$_\odot$ \citep{vanDokkum+2016}.

Reliable total mass measurements for statistical samples of UDGs are critical because we can use them to assess the degree to which these galaxies have underproduced stars, or ``failed", during their existence.
The first formation scenario for UDGs posited that the loss of gas at early times in dense environments led to massive failed galaxies in the Coma cluster \citep{vanDokkum+2015}. This suggestion was quickly followed up by an opposing model where it was proposed that UDGs represent the tail of high angular momentum halos, interpreting UDGs instead as ``puffy" low mass galaxies  \citep{Amorisco+2016}.  Subsequently, numerical simulations have introduced a number of other possible evolutionary factors \citep[for some examples see][]{DiCintio+2017,Chan+2018,Martin+2019,Wright+2021}. Although the situation is clearly more complex than suggested by early toy models, empirical estimates of M$_h$ are constraining for any scenario.

The other factor that is often cited as critical in UDG formation is the  environment \citep[e.g.,][]{saf,Chan+2018,Carleton+2019,Sales+2020,Wright+2021}. Interactions, either with individual galaxies or with the global environment, are often invoked to truncate star formation and produce the typically quiescent appearance of UDGs \citep{vanDokkum+2015,saf,Chan+2018,grishin}. The identification of UDGs in the field \citep[e.g.,][]{Martinez-Delgado+2016,Roman+2017,Leisman+2017,Greco+2018b} directly demonstrates that dense environments are not necessary for UDG formation. However, even more restrictive are subsequent observations that appear to show little if any change in the number of UDGs per total host environment mass across a wide range of environments \citep{vandenburg+2016,Roman+2017,vandenburg+2017,Karunakaran+2023,goto}. If environment plays a role beyond quenching star formation \citep[UDG color does depend on environment;][]{Prole+2019,Kadowaki+21}, creation and destruction of UDGs must be delicately balanced.

Our challenge is therefore four-fold. First, identify a sample of UDG candidates across environments in sufficient numbers for statistical study. This we do as described in Papers I-III and complete that task here (\S\ref{sec:data}, \ref{sec:processing}, and \ref{sec:catalog}). Second, estimate distances for as large a fraction of those candidates as possible to determine which candidates satisfy the UDG size criterion. We do this using the distance-by-association technique presented in Paper III (\S\ref{sec:distances}). As larger samples of spectroscopic redshifts become available the method will become more robust, so although we present our distance estimates we consider this to be a living catalog that will improve with time. Third, provide a measurement of the local environment (\S\ref{sec:environment}). Such an estimate is a byproduct of our distance estimation technique (\S\ref{sec:distances}). Finally, estimate M$_h$ for as large a sample as possible, which we do in \S\ref{sec:masses} using the technique presented by \cite{zb}. We close by discussing the implication of our mass measurements on the star formation efficiency of UDGs in \S\ref{sec:discussion}. We use a WMAP9 $\Lambda$CDM flat cosmology throughout with $\Omega = 0.287$, and H$_0 = 69.3$ km s$^{-1}$ Mpc$^{-1}$ \citep{hinshaw}. 
Magnitudes are on the AB system \citep{oke1,oke2}.

\section{The Data}
We report the results of our analysis of Data Release 9 (DR9) of the northern portion of the Legacy Survey, which includes observations obtained by the MOSAIC camera at the KPNO 4 m telescope (MzLS, Mayall z-band Legacy Survey) and the 90Prime camera \citep{90Prime} at the Steward Observatory 2.3 m telescope (BASS, Beijing-Arizona Sky Survey).  In addition to these telescopes, the full survey also employs DECam (DECaLS) at the CTIO 4m  that was the focus of our previous work described in Papers I-III. This paper presents our final SMUDGes catalog release and we do not intend to reprocess data using DR10 or any future release.

\begin{figure}[ht]
\begin{center}
\includegraphics[width=0.47\textwidth]{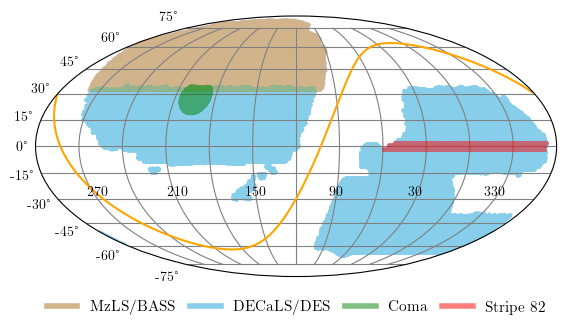}
\end{center}
\vskip -.5cm
\caption{Footprint of the sky covered in all three bands by the DR9 release of the Legacy Survey \citep{Dey+2019}. Observations used in this study from the northern part of the Legacy Survey are shown in tan (MzLS and BASS, jointly referred to as MB). Footprints of our previous work in Coma (Paper I), SDSS Stripe 82 (Paper II), and the full southern region of the Survey (DECaLS and DES) (Paper III) are displayed in green, red, and blue, respectively.  The Galactic plane is traced by the orange curve.}
\label{fig:footprint}
\end{figure}

Briefly, the Legacy Survey \citep{Dey+2019} was initiated to provide targets for the DESI survey drawn from  deep, three-band ($g=24.7$, $r=23.9$, and $z=23.0$ AB mag, 5$\sigma$ point-source limits) images. That survey covers about 14,000 deg$^2$ of sky visible from the northern hemisphere between declinations  approximately bounded by $-$18$^\circ$ and +84$^\circ$.  The footprint of DR9 (Figure \ref{fig:footprint}) also includes an additional 6,000 deg$^2$ extending down to $-$68$^\circ$ imaged at the CTIO by the Dark Energy Survey \citep{des}. As shown in the figure, the footprint of MzLS and BASS (hereafter jointly referred to as MB) has 3-band coverage of about 5,000 deg$^2$ at declinations $\gtrsim$32$^\circ$ with about 300 deg$^2$ overlapping the region observed by DECaLS.  

Because of the significant differences between the instrumentation used for MB  and DECaLS, we have modified the pipeline developed for DECam that was used in our earlier work.  Although the basic approach remains similar, there are noteworthy changes at various steps that we describe below.

\label{sec:data}
\section{Processing MB}
\label{sec:processing}

As previously mentioned, the northern Legacy Survey data comes from two different telescope/camera combinations. The MOSAIC camera (MzLS) provides imaging in the $z$ band and consists of four 4096$\times$4096 CCDs with a scale of 0.26\arcsec pixel$^{-1}$ and a field of view of $\sim$36\arcmin\  \citep{mosaic}\footnote{\href{https://noirlab.edu/science/sites/default/files/media/archives/documents/scidoc0111-en_0.pdf}{KPNO Mosaic-3 Imager User Manual}}.
Imaging in the $g$ and $r$ bands is obtained with the 90Prime camera (BASS), which contains four 4096$\times$4096 CCDs with a scale of 0.454\arcsec pixel$^{-1}$ and a field of view of 1.08$^\circ\times 1.03^\circ$ \citep{Zou}. 
BASS observations were taken from 12 November 2015 to 7 March 2019 and MzLS from 19 November 2015 to 12 February 2018\footnote{\href{https://www.legacysurvey.org/dr9/description/\#photometry}{www.legacysurvey.org/dr9/description/\#photometry}}.
As before, our processing and analysis of these data are performed on  the Puma
cluster at the University of Arizona High Performance Computing center\footnote{\href{https://public.confluence.arizona.edu/display/UAHPC/Resources}{public.confluence.arizona.edu/display/UAHPC/Resources}}. 

As in our previous work, we summarize the major steps in identifying potential UDGs as: 1) image processing that produces a preliminary list of candidates; 2) rejection of those that are likely to be Galactic cirrus contamination; 3) automated classification  screening for false positive detections; 4) visual confirmation of remaining candidates; 5) estimation of completeness, biases, and uncertainties using simulated sources; and 6) creation of the catalog. Details of these processes have been previously described and, other than brief summaries, we only address pipeline modifications here. 

In a significant departure from our previous methodology, we now use the 3-band coadded images generated by the Legacy Survey pipeline and publicly available on their website\footnote{\href{https://portal.nersc.gov/cfs/cosmo/data/legacysurvey/dr9/north/coadd/}{portal.nersc.gov/cfs/cosmo/data/legacysurvey/dr9/north/coadd/}} in our UDG search.  The Survey footprint is divided into 0.25 $\times$ 0.25 deg$^2$ ``bricks" as described by \cite{Dey+2019}.  During coaddition, the camera images are reprojected at a pixel scale of 0.262 arcsec with North up, converted into calibrated flux units of nanomaggies, and sky-subtracted.  The actual brick sizes are 3,600 $\times$ 3600 pixel$^2$, allowing a small amount of overlap.  We limit our analyses to those included in the survey-bricks-dr9-north.fits.gz file, which is included in the Legacy Survey's DR9.  This file has information for  each brick, including the median number of exposures in each filter. As in our earlier work, our automated classifier (Section \ref{subsec:classifier}) requires cutout images from all three bands and, therefore, we only include those bricks having at least one observation in each band.  We further exclude a small number of observations ($<$1\%) that were obtained in equatorial regions to cross-calibrate photometry with DECaLS and DES imaging \citep{Dey+2019}, leaving 83,038 bricks for analysis of which about 6\% overlap with DECaLS in the declination range of $\sim32^\circ -34^\circ$ (Figure \ref{fig:footprint}).

\subsection{Image Processing for MB}
\label{subsec:image processing}
After downloading brick images and supporting files from the Legacy Survey website we identify potential UDG candidates using the following steps: 
\begin{enumerate}
\item Although the coadded images have already passed through the Legacy Survey pipeline, we additionally replace saturated pixels with values obtained from neighboring pixels using the methodology described in Paper I.
\item We remove objects on bricks that are clearly too bright to qualify as UDG candidates.  As in Paper III, this is done using modeling to subtract sources that have SExtractor \citep{bertin} MU\_MAX values that are 2 mag arcsec$^{-2}$ brighter than a specified threshold in each band (24.0 for $g$, 23.6 for $r$, and 23.0 for $z$).
\item A crucial step in our detection pipeline is the use of wavelet transforms with tailored filters to isolate candidates of different angular scales.  When applied to all MB bricks, this results in a total of %\textbf{
46,374,815 detections, or an average of %\textbf{
$\sim$558/brick, the vast majority of which will not be classified as UDG candidates after further screening.

\item Spurious detections are limited by requiring that a potential candidate have  coincident detections (defined as  center-to-center separations $<$4\arcsec), in at least two of the three bands, with the resulting group of detections considered to be located at the mean centroid position. This requirement rejects all but %\textbf{
7,238,447
%} 
wavelet detections and results in %\textbf{
3,521,143
%} 
separate groupings with an average of %\textbf{
$\sim$ 42 groups/brick.

\item
At this point in our pipeline the vast majority of
candidates will not survive further screening and, as described in our earlier work, we limit the number of detections requiring time-consuming GALFIT \citep{peng} modeling by obtaining
much faster, rough parameter estimates using the
LEASTSQ function from the Python SciPy library \citep{jones}. Other than modeling detections on bricks rather than CCDs our approach is
unchanged from Paper III. Because this is only used as a coarse
screen, we fit an exponential S\'ersic model (n=1)
to each candidate on a brick and require that the
results meet conservative parameter thresholds of
$r_e \ge 4^{\prime\prime}$ and $\mu_0 \ge$  23.0, 22.0
and 21.5 mag arcsec$^{-2}$ for $g$, $r$, and $z$, respectively.
In a departure from our previous work, we only
require that a candidate successfully meet these
criteria in one of the 3 bands comprising a brick,
leaving a total of 
624,469 detections comprising
499,866 distinct candidates that survive this step. Note that while changes such as this one will result in differences in the candidate UDGs between the southern and northern regions, some differences were unavoidable given the data quality differences. Nevertheless, we attempt to homogenize the catalog by evaluating measurement biases and completeness separately for the southern and northern surveys.
\item We perform an initial GALFIT screen of each candidate using a fixed S\'ersic index of $n$ = 1, without incorporating the point spread function (PSF) into the model.  In our prior work, we allowed GALFIT to generate its own sigma image, but we now provide one using the publicly available inverse variance images created by the Legacy Survey pipeline during coaddition. As in Paper III, we use generous acceptance thresholds of $r_e \ge 4\arcsec$, $b/a \ge 0.34$,  and $\mu_{0,g} \ge$ 23 mag arcsec$^{-2}$ or $\mu_{0,z} \ge$ 22 mag arcsec$^{-2}$ if there is no available measurement of $\mu_{0,g}$.  A total of 
%\textbf{
86,746 candidates meet these criteria. Important details, such as masking, are described fully in Papers I, II and III. Here we only note that the masking removes nearby objects, including any bright source at the center of the UDG candidate, to avoid the influence of possible nuclear star clusters in the model fitting. The central surface brightness that we use throughout our selection is that inferred from the fitted model, not one directly measured at the source center.

\item  Our final image processing step uses GALFIT with a variable S\'ersic index and an estimate of the PSF to model the remaining candidates as described in Paper III.  We again use inverse variance images provided by the Legacy Survey to create the sigma image. In another departure from our earlier pipeline we now use their PSF images to estimate the PSF by taking the median value of a 7$\times$7 pixel region centered on the candidate.

\end{enumerate}

When we compare the initial results from our pipeline to matched DECaLS candidates in the overlapping region, we find large biases in the structural parameters and $z$-band photometry with much smaller differences for photometric estimates measured in the $g$ and $r$ bands. There are no significant biases when we reprocess these same candidates using bricks from DECaLS, indicating that the problem is not pipeline-related. Other than for $r_e \ge 10$\arcsec, where values using bricks are $\sim 25 \%$ less than those of CCDs, there are no significant biases in either structural or photometric estimates when we reprocess these same candidates using bricks from DECaLS.  While some discrepancies, which seem to be limited to estimates of $r_e$ at large effective radii, may be attributable to changes in the pipelines, the vast majority of the differences appear to be related to the processing of the MB images.   Moreover, when we reprocess images from single filters using bricks, discrepancies for both $r_e$ and photometric properties are isolated to the $z$-band and become more significant for the larger candidates. We attribute the behavior we find to oversubtraction of the background in DR9 processing of the $z$-band images of large, low surface brightness systems\footnote{This issue is known among the Legacy Survey team and thought to have originated from revisions of the processing pipeline for DR9. It is being addressed (priv. comm.)]}.  The discrepancies in morphological parameters significantly improve when the $z$-band is omitted from the stack. Therefore, we proceed without the $z$-band images in the stack.  However, even with this modification estimated effective radii are slightly smaller than those found in DECaLS.
Fewer than 10\% of our candidates have $r_e$ $\ge $10\arcsec\ and we do correct for biases using our artificial source simulations, so we do not expect this modest discrepancy to affect our conclusions. 
This comparison highlights the problems inherent in comparing results from different telescopes using different pipelines; especially if there is no avenue for assessing relative biases.  

After applying our final criteria of $r_e \ge 5.3\arcsec$, $\mu_{0,g}\ge$ 24 mag arcsec$^{-2}$ (or $\mu_{0,z}\ge$ 23 mag arcsec$^{-2}$ if GALFIT failed to model $g$), $b/a \ge $ 0.37, and $n < 2$ we are left with 
%\textbf{
22866 candidates available for further evaluation.

\begin{deluxetable*}{lrrr}
\tablecaption{Number of detections and UDG candidates in MB vs. processing step}
\label{tab:screening}
\tablewidth{0pt}
\tablehead{
\colhead{Process}&Description in Text&
\colhead{Detections} &
\colhead{UDG Candidates}\\
}
\startdata
Wavelet screening & \S3.1, Step 3 & 46,374,815  & NA \\
Object matching &\S3.1, Step 4  & 7,238,447 & 3,521,143\\
S\'ersic screening &\S3.1, Step 5 & 624,469  & 499,866\\
Initial GALFIT screening & \S3.1, Step 6& NA & 86,746 \\
Final GALFIT screening & \S3.1, Step 7& NA & 22,866  \\
Cirrus screening & \S3.2 & NA & 8,478  \\
Duplicate removal  & \S3.3& NA & 8,219\\
Automated classification & \S3.4& NA & 1,415\\
Color criterion & \S3.4& NA & 1,374\\
Visual Examination & \S3.5& NA & 1,269\\
\enddata
\end{deluxetable*}

\subsection{Screening of Spurious Sources Caused by Cirrus}
\label{subsec:cirrus}
\begin{figure}[ht]
\begin{center}
\includegraphics[width=0.47\textwidth]{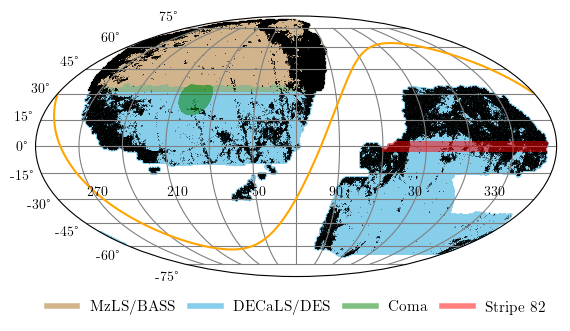}
\end{center}
\vskip -.5cm
\caption{Cirrus contamination within DR9 of the Legacy Survey footprint. Regions in black exceed our dust proxy thresholds of either 0.1 MJy/sr for WISE 12 $\mu$m map or 0.05 for  $\tau_{353}$ and comprise $\sim$28\% of the entire footprint. The Galactic plane is shown in orange.}
\label{fig:dust}
\end{figure}

Large regions of the Legacy Survey footprint are contaminated by Galactic cirrus that can result in spurious detections that may be difficult to differentiate from legitimate UDG candidates (Figure \ref{fig:dust}).  As in Paper III, we address this problem by rejecting as probable dust any candidates having single point values exceeding 0.05 in the $\tau_{353}$ dust map \citep{planck} or 0.1 MJy/sr in the WISE 12 $\mu$m map \citep{meisner}. Using these criteria, we found in Paper III that 1.7\% of the Coma region, $\sim$43\% of the Stripe 82 footprint, and $\sim$31\% of the entire DECaLS footprint  exceeds these thresholds.  As shown in Figure \ref{fig:dust},  about 22\% of MB and 28\% of the entire DR9 footprint are contaminated with cirrus.  After applying the dust criteria to our candidates, a total of 
%\textbf{
14388 are rejected, leaving 
%\textbf{
8478 in the MB footprint for further analyses.

\subsection{Screening of Duplicates}
\label{subsec:duplicates}

Before further processing, we eliminate duplicate entries, which we define as candidates lying within 10\arcsec\ of each other.  We previously accepted the first entry of a group with duplicates and rejected the remainder.  We now select the one with the smallest separation from the center of the cutout and reject the others on the assumption that the candidate with coordinates closest to those provided to GALFIT is the one most likely to be the desired entry.  Because it is possible that detections separated by 10\arcsec\ could each represent a legitimate, distinct candidate, we visually inspected all 
%\textbf{
10 cases where separations were between 5\arcsec\ and 10\arcsec\ and found none that contained {\sl bona fide} separate candidates.  Most of the closely spaced duplicates result from detections of the same object in different bricks with one or more lying in an overlap region outside of the main 0.25 $\times$ 0.25 deg$^2$ brick area. Other causes include residual artifacts, tidal material, cirrus not rejected by our dust criteria, background clusters, and in a few cases, large, probably nearby, candidates that were detected multiple times during wavelet filtering. Our criterion eliminates 
%\textbf{
259 potential candidates, leaving %\textbf{
8219 for further classification. 

\subsection{Automated Classification}
\label{subsec:classifier}
Our approach to computer classification is described in detail in the appendix of Paper I with modifications addressed in Paper II.  Briefly, we use a convolutional neural network which was trained on visually classified cutouts downloaded from the Legacy Survey in the Stripe 82 and Coma regions.  We make no changes for the current study and use the prior trained network and weights for classification, resulting in 
%\textbf{
1415 of the 
%\textbf{
8219 being designated as UDG candidates.  As explained in Paper II, based on visual inspection we find that objects with $g-r$ colors $>$1.0 mag are unlikely to be UDGs and so reject 
%\textbf{
41 such objects from further consideration.

\subsection{Visual Confirmation}
\label{subsec:confirm}
As part of the development of our automated classifier in Paper II we found from visual examination that about 2.6\% (8/306) of the candidates identified as potential UDGs in a test set were false positives.  Paper III had similar results with about 2.8\% (162/5760) being visually classified as false positives.  We again wish to  minimize the effects of false positives in our current catalog and DZ and RD visually review the
%\textbf{
1374 remaining candidates in MB. Each reviewer initially classifies each candidate as a potential UDG, a false positive, or questionable. Those with disagreements or labeled as questionable are again classified by both reviewers.  This procedure results in both reviewers labeling %\textbf{
1269 as UDG candidates and 
%\textbf{
80 as false positives with disagreement on %\text{
25. To minimize the number of false or ambiguous detections in our catalog, we consider any disagreements to be false positives resulting in 
%\textbf{
105/1374 (7.6\%) being labeled as such.  This fraction is almost triple those found in Papers II and III.  We attribute this rise to our use of the training set drawn from stacked images obtained from DECaLS and DES, which are much deeper than the images used in the current study.\footnote{\href{https://www.legacysurvey.org/status/}{www.legacysurvey.org/status/}}  False positives are mitigated because all candidates are visually confirmed. 

\begin{figure}[ht]
\begin{center}
\includegraphics[width=0.47\textwidth]{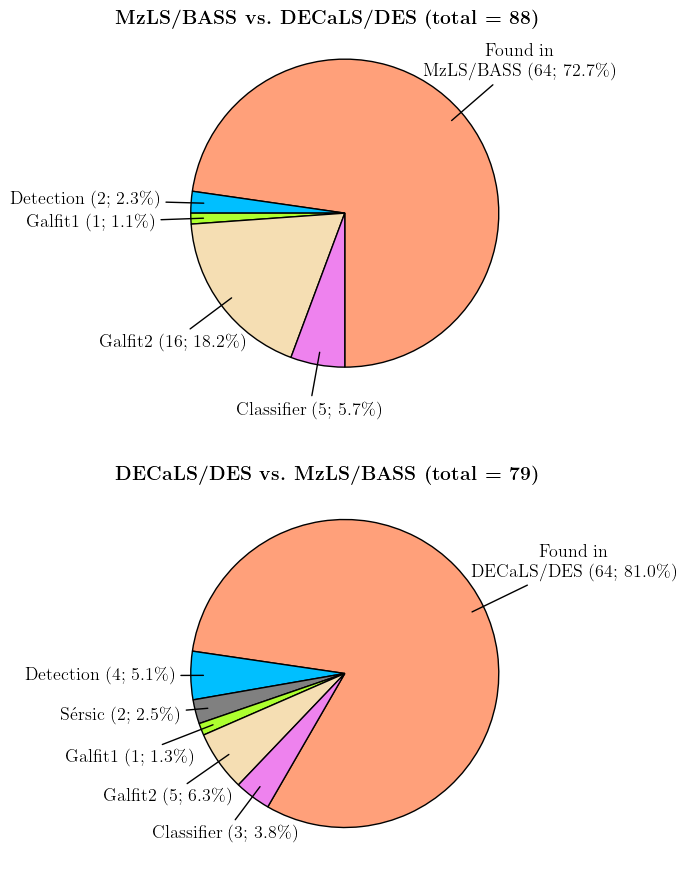}
\end{center}
\vskip -.5cm
\caption{Top panel: Candidates found or missed in MzLS/BASS that are present in the DECaLS/DES catalog. Bottom panel: Candidates found or missed in DECaLS/DES that are present in the MzLS/BASS catalog.  Detection: Inadequate or no wavelet detections (\S\ref{subsec:image processing}, Step 4); S\'ersic: Failed to meet our S\'ersic screening criteria (\S\ref{subsec:image processing}, Step 6); Galfit1: Failed our initial GALFIT screen (\S\ref{subsec:image processing}, Step 8); Galfit2: Failed our final GALFIT estimates (\S\ref{subsec:image processing}, Step 9); Classifier: Failed to meet the probability threshold required by our automated classifier (\S\ref{subsec:classifier}).}
\label{fig:discrepancies}
\end{figure}

To the extent that our simulations (\S \ref{subsec:simulations}) accurately represent actual UDGs, our estimates of completeness and bias should help compensate for differences in false negatives between the MB and DECaLS.  Nonetheless, we attempt to quantify some of the causes of false negatives by investigating outcomes in the $\sim$300 deg$^2$ region that overlaps both surveys at their abutting edges (Figure \ref{fig:footprint}) with results shown in Figure \ref{fig:discrepancies}. A total of %\textbf{
79 candidates from the northern and %\textbf{
88 from the southern contributions to this region made it through our entire pipeline, including visual confirmation. Of these, 64 were common to both surveys, leaving %\textbf{
15 (19\%) in the northern portion and %\textbf{
24 (27\%) in the southern portion unmatched.  

Only a few of the discrepancies resulted from undetected candidates and none because of visual confirmation.  Most of the candidates not identified by the MzLS/BASS pipeline failed to meet our UDG criteria during final GALFIT modeling (Step 7 of Section \ref{subsec:image processing}). Although %\textbf{
13/16 had $r_e\ge$ 4\arcsec, these did not meet our $r_e=5.3$\arcsec\ threshold and this was the most common cause of failure. This is not surprising since, as discussed  (\S\ref{subsec:image processing}), GALFIT estimates of $r_e$ for the northern survey tend to be smaller than those in the southern survey. Although our correction for this bias compensates for the difference in those surviving our entire pipeline, we still reject those that do not meet our UDG criteria at that point in the processing pipeline.  About half of the candidates rejected by our automated classifier were close ($>$0.9) to the threshold (0.99) that we use for accepting a candidate (Paper II).  Two candidates in DECaLS were rejected during S\'ersic screening, likely because the individual CCD images processed in that study are noisier than the coadded bricks used in MzLS/BASS. This discussion highlights how sensitive membership is to the fine details of the data quality even when the analysis is done consistently, particularly near the limits of the selection criteria. We should not be surprised that a number of UDGs appear or disappear among different overlapping surveys because of different selection algorithms. We described similar results in Papers II and III, where we presented in more detail a comparison among catalogs from independent surveys. We recover most candidates in those surveys that satisfy the SMUDGes selection criteria.
For example, when comparing to an H{\small I}-selected sample \citep{Leisman+2017}, we recover 39 of 41 sources that match our selection criteria.
Although in general we do find a few discrepancies, we conclude that there is no evidence for systematic biases among the surveys.

\subsection{Estimating completeness, biases, and uncertainties using simulated UDGs}
\label{subsec:simulations}
In Papers II and III we estimated uncertainties and recovery completeness by planting simulated UDGs at random locations and separately processing them with the same pipeline as used for our real sources, including automated classification. 
Detailed descriptions and rationales for our approach to modeling are presented in Paper II with minor modifications and limitations discussed in Paper III and will not be repeated here except when needed for clarification.  We continue to use S\'ersic profiles with random structural and photometric properties, but because we now have a smaller footprint, we increase the initial simulation density from 600 to 1800 deg$^{-2}$ ($\sim$112 per brick). We further avoid brick overlap regions by restricting the central locations to only those pixels within the 0.25 $\times$ 0.25 deg$^2$ defined by the brick. Because we require a separation of at least 40\arcsec\ between simulations the total number created is %\textbf{
7,295,356 for an average of about 88 per brick. 

Our basic approach to modeling prevents us from including the full range of simulation parameter space. We mitigate the effects of this problem by expanding the thresholds used for our science candidates and accepting simulations with $22.5 <\mu_{0,g}< 27.5$ mag arcsec$^{-2}$, $3.5 <r_e < 20$\arcsec, $b/a > 0.25$, and $0.1 <n < 2$ that also meet our dust and color criteria with 
%\textbf{
1,098,760 passing automated classification\footnote{The brighter surface brightness limit was incorrectly quoted as 23.5 mag arcsec$^{-2}$ in Paper III. In practice, it was the same as that used here.}. 

As described in Paper  II, all models use the four parameters that we consider the most appropriate for exploring the variable under consideration.  Completeness and uncertainty estimates are obtained using polynomial models created with the PolynomialFeatures function from the Python Scikit-learn library \citep{sklearn} and a four layer neural network implemented with Keras \citep{keras}.

\subsubsection{Uncertainties}
Parameter uncertainties for simulated sources are defined as the difference between final GALFIT values and the values used when creating the associated simulations (GALFIT $-$ input). Because uncertainties are generally asymmetric, we define the bias as the median difference and the ``1$\sigma$" confidence limits as the 15.1 and 84.9 percentiles of the distribution for a set of similar simulated objects. 

Because polynomial models, especially of high order, may extrapolate very poorly for data points lying outside of the fitted range, our simulation range for individual parameters extends beyond those expected for our science targets.  As in Paper III, we further mitigate this problem by using 2\textsuperscript{nd} degree polynomial models to fit the simulation data. We continue to set all position angle, $\theta$, biases to zero in the catalog because these values are negligible and we want to avoid adding noise.

\subsubsection{Completeness}
\label{completeness}
We define completeness as the probability that a candidate with given structural and photometric parameters will survive our entire pipeline.  This is assessed using four modeled parameters ($\mu_{0,g}$, $r_e$, $b/a$, and $n$) and again uses a 2\textsuperscript{nd} degree polynomial to fit the simulation results.  We apply bias corrections to our catalog entries before estimating their completeness probabilities.  Completeness for very large candidates, primarily in Virgo, was a problem in Paper III because our model $r_e$ only extend to 20\arcsec; however, no true candidate in the current study reaches that threshold so this issue is not a concern here.

\section{The Catalog}
\label{sec:catalog}
As noted in \S\ref{subsec:confirm} there are 64 candidates common to DECaLS and MB.  We omit these from the MB sample to avoid duplicate entries in the final merged catalog, leaving a total of 1310 new entries. Because we want to keep our simulation pipeline, which did not include visual examination, identical to our science pipeline, we retain in the catalog, but flag, candidates that we visually identified as false positives.  These should be omitted from any conclusions drawn from our results.  Descriptions of the catalog entries are presented in Table \ref{tab:catalog} and the full catalog is available in the electronic version. Each parameter entry includes its GALFIT estimate as well as its bias and confidence limits produced by our models.  Any entry which required extrapolation of the fitted model beyond the range of the constraints is flagged and should be used with caution (\S3.6). Users should apply the bias values by subtracting those presented in the catalog from the corresponding uncorrected measurements when drawing conclusions from the data. 

As mentioned in  \S\ref{subsec:image processing}, because of systematic problems in the northern $z$-band stacked images the magnitudes in MB tend to be higher than those in DECaLS.  This offset is 0.5 mag and essentially independent of the estimated magnitude.  Entries in the catalog are corrected for this bias by subtracting 0.5 mag from the GALFIT estimates for mag$_{z}$ and $\mu_{0,z}$ is recalculated from these new values and the structural parameters assuming a S\'ersic profile.  Uncertainties and completeness are estimated from these revised entries.  Nonetheless, all $z$-band entries for MB sources should be considered suspect and while they may be adequate for statistical conclusions, individual entries should be used with caution.

Parameters are corrected for bias before their completeness values are estimated. Completeness estimates may be suspect (Comp\_flag $\ne 0$) for either of two reasons.  The parameters may be outside of the parameter space defined by our completeness model and these have flag = 1.  Alternatively, the bias correction derived from the uncertainty model may be unreliable and these have flag = 2.  In either case, the results should be used with caution.

Photometric parameters are not corrected for extinction, but extinction values are included in the catalog for those wishing to use them. Our extinction estimates (A$_g$, A$_r$, A$_z$) are calculated using the SDSS $g$, $r$, and $z$ Legacy Survey extinction coefficients\footnote{\href{https://www.legacysurvey.org/dr9/catalogs/\#galactic-extinction-coefficients}{legacysurvey.org/dr9/catalogs/\#galactic-extinction-coefficients}} with E(B-V)$_{SFD}$ estimated using the dustmaps.py \citep{green} SFD dust map based on the work of \cite{SFD}. 

\begin{figure*}[h]
\begin{center}
\includegraphics[width=0.9\textwidth]{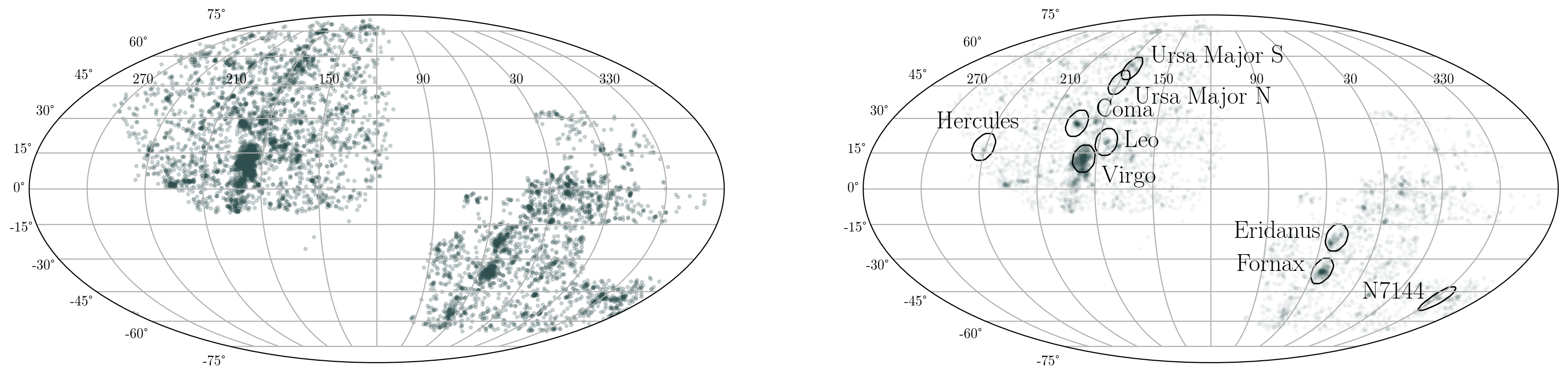}
\end{center}
%\vskip -.5cm
\caption{The distribution of candidate UDGs across the sky in right ascension and declination. The right panel is the same as the left except that we have decreased the opacity of each point to help highlight the higher density regions. We label some well known local overdensities for reference.} 
\label{fig:sky_dist}
\end{figure*}

We recommend that images be reviewed in any study drawing conclusions based on individual candidates, particularly if those are extreme in any way (e.g., largest, faintest, etc.).
The sky distribution of the merged SMUDGes catalog is shown in Figure \ref{fig:sky_dist}. From now on, we discuss the merged southern and northern candidate sample.

\begin{deluxetable*}{lrr}
\caption{The Complete Catalog$^a$}
\label{tab:catalog}
\tablehead{
\colhead{Column Name}&
\colhead{Description}&
\colhead{Format}
}
\startdata
SMDG&Object Name&SMDG designator plus coordinates\\
RA&Right Ascension (J2000.0)&decimal degrees\\
Dec&Declination (J2000.0)&decimal degrees\\
r\_e&effective radius&angular (arcsec)\\
r\_e\_upper\_uncertainty&effective radius 1$\sigma$ upper uncertainty&angular (arcsec)\\
r\_e\_bias&effective radius measurement bias&angular (arcsec)\\
r\_e\_lower\_uncertainty&effective radius  1$\sigma$ lower uncertainty&angular (arcsec)\\
r\_e\_flag&effective radius uncertainty model flag&0 = good, 1 = extrapolated\\
AR&axis ratio ($b/a$)&unitless\\
AR\_upper\_uncertainty&axis ratio 1$\sigma$ upper uncertainty&unitless\\
AR\_bias&axis ratio measurement bias&unitless\\
AR\_lower\_uncertainty&axis ratio 1$\sigma$ lower uncertainty&unitless\\
AR\_flag&axis ratio uncertainty model flag&0 = good, 1 = extrapolated\\
n&S\'ersic index&unitless\\
n\_upper\_uncertainty&S\'ersic index 1$\sigma$ upper uncertainty&unitless\\
n\_bias&S\'ersic index measurement bias&unitless\\
n\_lower\_uncertainty&S\'ersic index 1$\sigma$ lower uncertainty&unitless\\
n\_flag&S\'ersic index uncertainty model flag&0 = good, 1 = extrapolated\\
PA&major axis position angle&defined to be [$-$90,90) measured \\
&&N to E, in degrees\\
PA\_upper\_uncertainty&major axis position angle 1$\sigma$ upper uncertainty&degrees\\
PA\_bias&major axis position angle measurement bias&degrees\\
PA\_lower\_uncertainty&major axis position angle 1$\sigma$ lower uncertainty&degrees\\
PA\_flag&major axis position angle uncertainty model flag&0 = good, 1 = extrapolated\\
mu0\_$X$&central surface brightness in band $X$ ($X \equiv$ g,r,z)&AB mag arcsec$^2$\\
mu0\_$X$\_upper\_uncertainty&central surface brightness 1$\sigma$ upper uncertainty in band $X$&AB mag arcsec$^2$\\
mu0\_$X$\_bias& central surface brightness measurement bias in band $X$&AB mag arcsec$^2$\\
mu0\_$X$\_lower\_uncertainty&central surface brightness 1$\sigma$ lower uncertainty in band $X$&AB mag arcsec$^2$\\
mu0\_$X$\_flag&central surface brightness uncertainty model flag in band $X$&0 = good, 1 = extrapolated\\
mag\_$X$&total apparent magnitude in band $X$&AB mag\\
mag\_$X$\_upper\_uncertainty&total apparent magnitude 1$\sigma$ upper uncertainty in band $X$&AB mag\\
mag\_$X$\_bias&total apparent magnitude measurement bias in band $X$&AB mag\\
mag\_$X$\_lower\_uncertainty&total apparent magnitude 1$\sigma$ lower uncertainty in band $X$&AB mag\\
mag\_$X$\_flag&total apparent magnitude uncertainty model flag in band $X$&0 = good, 1 = extrapolated\\
Rejected&rejected based on visual inspection&0 = good, 1 = rejected,\\
&&2 = observers disagreed\\
SFD&Optical depth at SMDG location from \cite{SFD}&unitless\\
A\_$X$&Corresponding extinction at SMDG location in band $X$&AB mag\\
Comp&Fractional completeness for similar UDGs&unitless\\
Comp\_flag&Completeness model flag&0 = good, 1=extrapolated, \\
&&2=biases extrapolated\\
cz&recessional velocity& km s$^{-1}$\\
cz\_type&redshift source&line of sight overdensity (OverDen),\\
&& cluster member (specific cluster name),\\
&& spectroscopic (specz)\\
sigma\_est&estimated  internal velocity dispersion&km s$^{-1}$\\
mass\_h\_est&estimate of log(M$_h$)&log(M/M$_\odot$)\\
env\_sigma&$\sigma_v$ of environment &km s$^{-1}$\\
env\_n&number of galaxies in environment&unitless\\
source&DECaLS (D) or MzLS/BASS (MB)&unitless\\
\enddata
\tablenote{The catalog is available as the electronic version of this Table.}
\end{deluxetable*}

\subsection{Estimating Distances}
\label{sec:distances}

Because UDGs are defined in terms of physical size, distance estimates are essential. This presents a challenge for our sample because it is not exclusive to rich galaxy clusters and groups. 
To estimate distances, we 
apply the distance estimation technique based on line-of-sight overdensities presented in Paper III. Additionally, we also assign those candidates projected  onto the Coma, Fornax, and Virgo clusters, the richest clusters in our survey (Figure \ref{fig:sky_dist}),
the corresponding distance of the associated cluster. To reprise the former, we examine the distribution of galaxies with known redshifts around the line of sight to each UDG candidate, searching for possible overdensities with which to associate the candidate. We only include galaxies that are projected within 1.5 Mpc of the UDG candidate as calculated using the redshift of the galaxy in question. We assess whether the association with an overdensity is unambiguous using both a set of fixed criteria and machine learning.  These criteria are described in detail in Paper III and result in the exclusion of any line of sight with multiple, independent overdensities that create ambiguity in the assigned redshift. We exclude objects with estimated $cz < 1800$ km s$^{-1}$ so that we can reliably estimate the distance using its Hubble velocity. In Paper III we demonstrated that the technique has an accuracy rate of roughly 70\% where accuracy is defined to mean that $\Delta z < 3 \sigma_z$ ($\sigma_z$ is our estimate of the  redshift uncertainty and corresponds to  the velocity width of the associated line of sight galaxy grouping). 

\begin{figure}[ht]
\begin{center}
\includegraphics[width=0.45\textwidth]{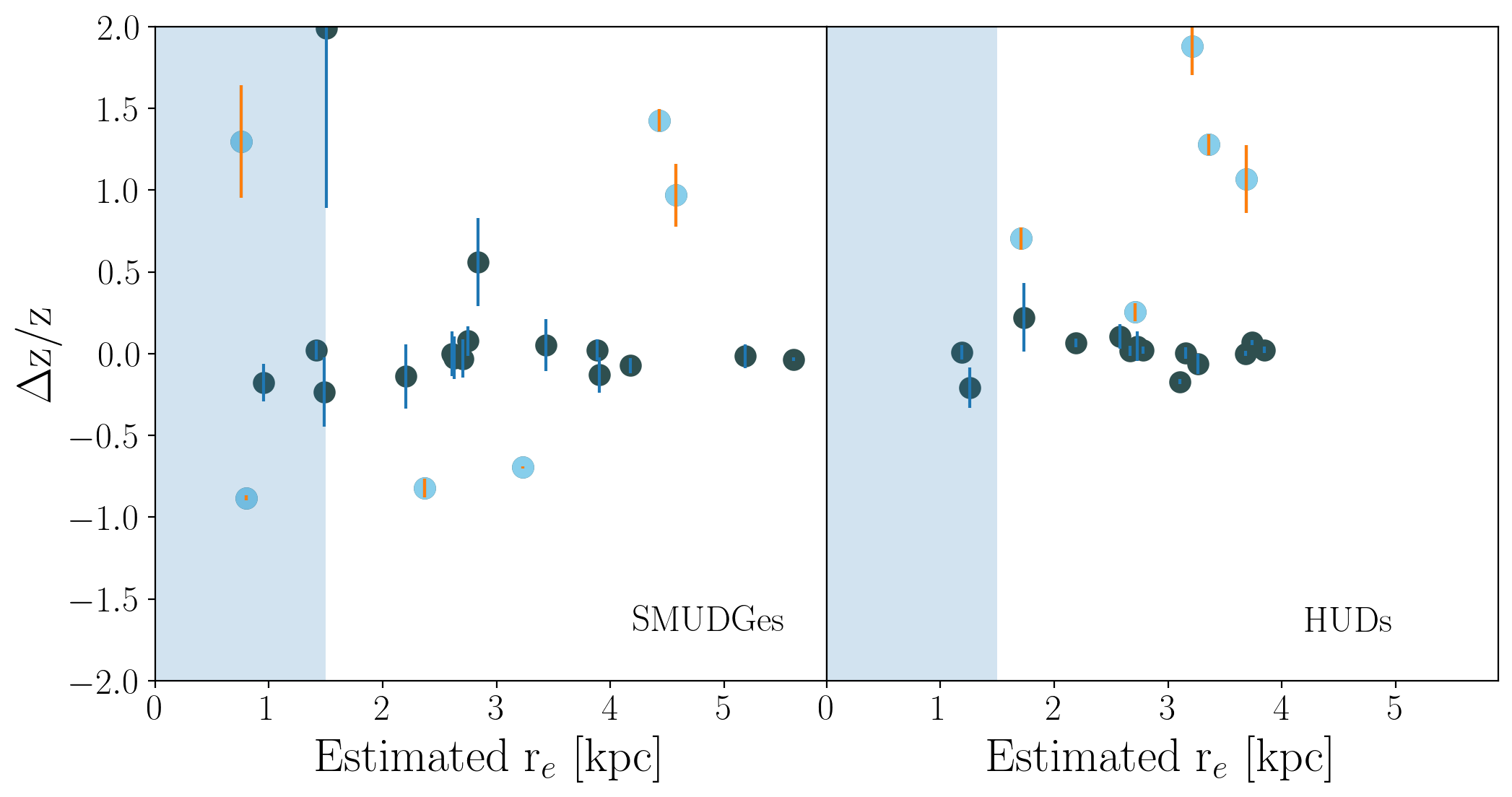}
\end{center}
\vskip -.5cm
\caption{Fractional redshift error vs. physical size. We plot the difference between the estimated and spectroscopically measured redshifts, divided by the redshift, as a function of inferred candidate size. The darker points represent candidates for which the estimate was within 3$\sigma$ of the measured value, while those in the lighter symbols those for which it was not. The left panel contains results for SMUDGes candidates with spectroscopic redshifts from the \cite{Kadowaki+21} compilation and the right panel for H{\small I}-selected ultra-diffuse galaxies (HUDs) from \cite{Leisman+2017}. The shaded area highlights the region where the candidates are no longer classified as UDGs due to their small physical size.} 
\label{fig:z_est}
\end{figure}

Using the complete catalog now available, and a few more UDGs with spectroscopic redshifts, we confirm (Figure \ref{fig:z_est}) that this technique results in an accuracy of $\sim$ 70\% when examining the SMUDGes candidates with measured redshifts \citep{Kadowaki+21}. The accuracy is 74\% when examining the sample of H{\small I}-selected low surface brightness galaxies from \cite{Leisman+2017}.  The results obtained for the latter sample show that the estimated redshift accuracy does not change appreciably when considering H{\small I}-rich galaxies, which presumably lie in lower density regions. However, in either case the redshift estimate yield is $<$ 20\% (18 and 13\% respectively), suggesting that we will only be able to estimate redshifts for a small fraction of the SMUDGes catalog using this technique and that the resulting sample may be skewed somewhat to denser than average environments.

To complement this set of estimated redshifts, we  assign any candidate without an estimated redshift and projected within 1, 1.5, and 2.3 Mpc from the Fornax, Virgo, and Coma clusters, respectively, the redshift of the host cluster. 
Finally, we include those candidates with spectroscopic redshifts (replacing any estimated redshift with the spectroscopic one). 

\begin{figure}[ht]
\begin{center}
\includegraphics[width=0.45\textwidth]{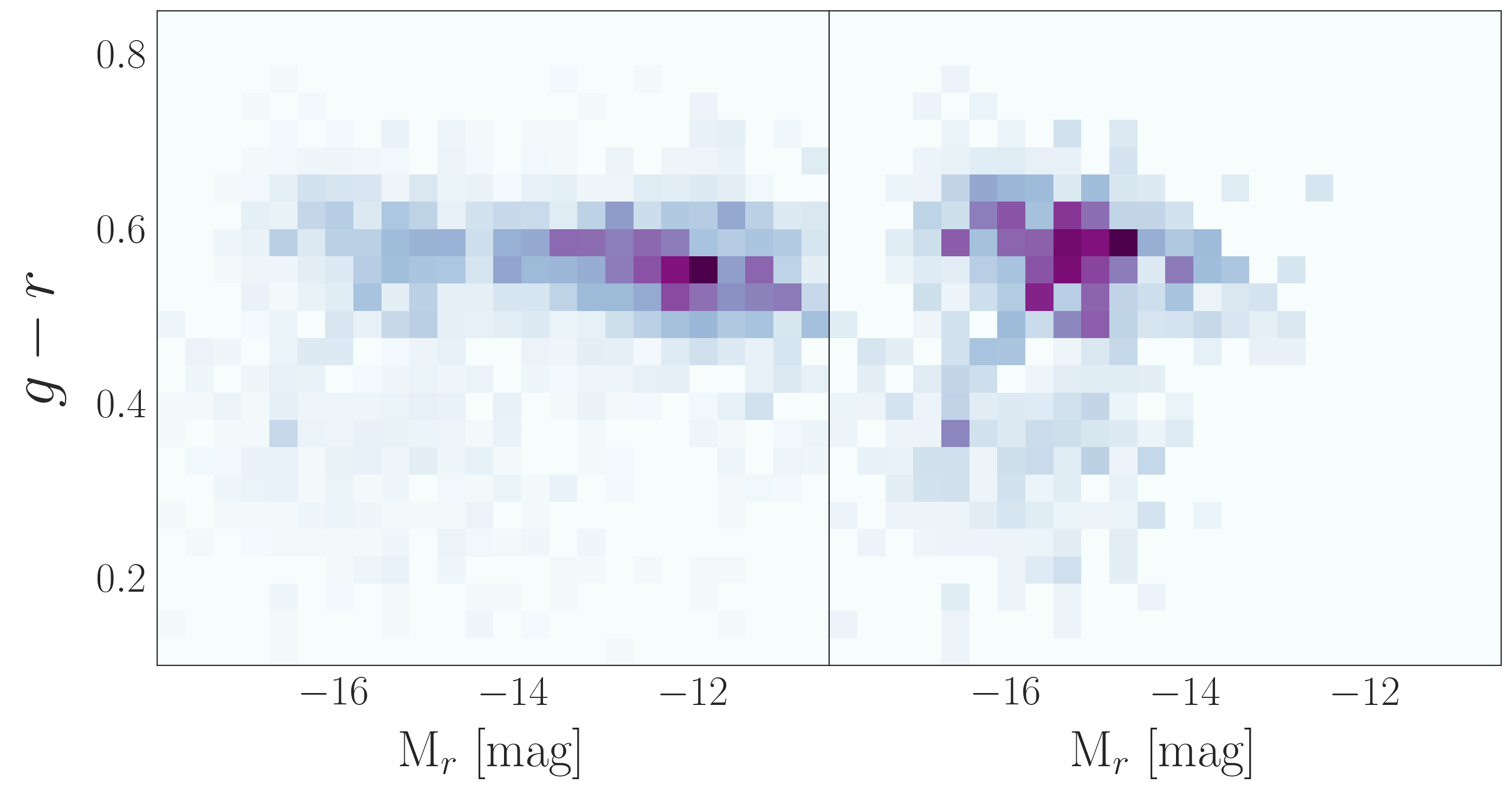}
\end{center}
\vskip -.5cm
\caption{Color-magnitude distribution for SMUDGes candidates with estimated redshifts. In the left panel we include all candidates with $cz_{est}$ and in the right panel only those that at the estimated distance have $r_e \ge 1.5$ kpc. The intensity scaling of each bin corresponds to a linear scaling of the number of objects in the bin, with different normalization in the two panels.}
\label{fig:color_mag}
\end{figure}

In total, we present redshift estimates for 1525 candidates in the catalog and confirm as UDGs ($r_e \ge 1.5$ kpc) 585 candidates. This UDG fraction is not expected to be representative for the survey because candidates in Virgo and Fornax, which are numerous and nearby, are far less likely to satisfy the physical $r_e$ criterion given our angular size selection. The properties of these systems is presented in Figure \ref{fig:color_mag}. For the bulk of our candidates we are unable to recover an estimated $cz$ and validate them as UDGs. A priority for future research in this area must be expanding the reach of redshift estimation techniques. Nevertheless, there are alternative approaches, such as correlation analyses, that can bring the power of the full catalog to bear on certain questions \citep[][]{Prole+2019,greene,goto}.
For the remainder of our discussion, we limit ourselves to the SMUDGes candidates with redshift estimates. Our estimates for the recessional velocities, $cz_{est}$, are included in the catalog, but they are likely to change in the future as the training sample and methodology improve. 

As with the full catalog itself, we caution that in selecting samples of unusual objects the likelihood of catastrophic failure becomes greater. For example, when selecting large ($r_e \ge 4$ kpc), blue galaxies, the $cz_{est}$ failure rate is closer to 50\% (Table \ref{tab:blues}). The increase in the failure rate arises because blue galaxies are less likely to be physically associated with clear galaxy overdensities and selecting the largest UDGs, which are rare, is likely to select for candidates with grossly overestimated values of $cz_{est}$. This overestimation is in fact the case for the three catastrophic failures (SMDG0031454+024256, SMDG0200102+284950, and SMDG0806124+153015).

\begin{deluxetable}{lrr}
\tablecaption{Redshift Comparison for Large Blue UDGs} 
\label{tab:blues}
\tablewidth{0pt}
\tablehead{
\colhead{Name}&
\colhead{Spectroscopic $cz$} &
\colhead{Estimated $cz$}\\
\colhead{}&
\colhead{[km s$^{-1}$]}&
\colhead{[km s$^{-1}$]}\\
}
\startdata
SMDG0031454+024256 & 2377 & 4683\\
SMDG0200102+284950 & 168 & 4874\\
SMDG0803340+090730 & 4412 & 4619\\
SMDG0806124+153015 & 1979 & 4803\\
SMDG0915558+295527 & 7234 & 6711\\
SMDG1601538+162909 & 10626 & 10464\\
\enddata
\end{deluxetable}

\subsection{Estimating Environment}
\label{sec:environment}

\begin{figure}[ht]
\begin{center}
\includegraphics[width=0.45\textwidth]{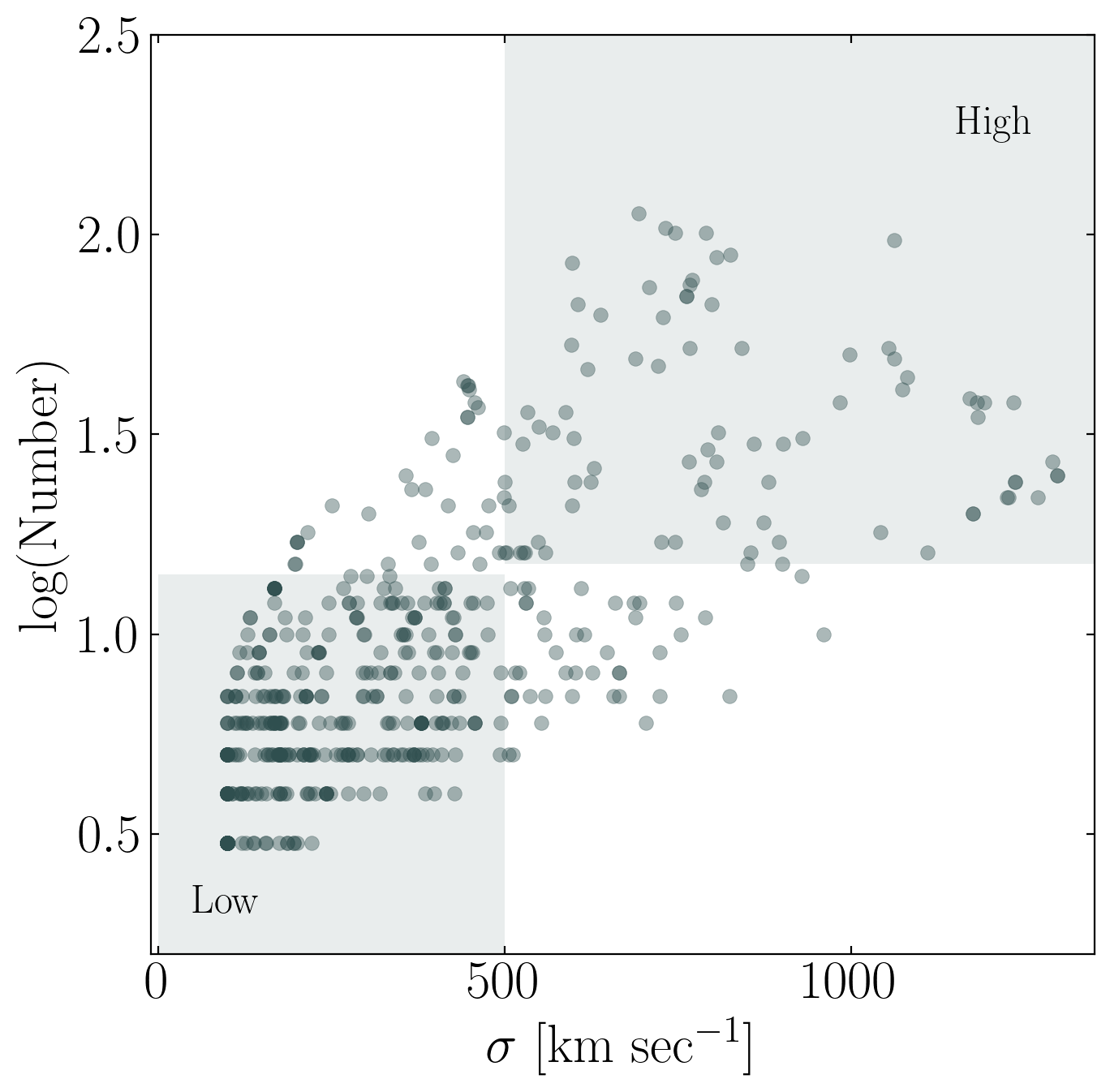}
\end{center}
\vskip -.5cm
\caption{Comparison of two environmental tracers. The velocity dispersion, $\sigma_v$, and number of galaxies  refers to the parameters of the overdensity associated with an individual UDG candidate. We show the results for all lines-of-sight, regardless of whether we accepted the estimated redshift. The two quantities track each other as expected, although with large scatter. Because of the large scatter, we choose to define low density environment as satisfying both $\sigma_v \le 500$ km s$^{-1}$ and $N < 15$, and high density environment as satisfying both $\sigma_v > 500$ km s$^{-1}$ and $N \ge 15$. The selected regions are shaded and labeled.}
\label{fig:environment}
\end{figure}

As a byproduct of the distance estimation technique we have measurements of both the velocity dispersion of the associated overdensity and the number of galaxies in the overdensity. Both of these measurements have significant potential problems. We derive the velocity dispersion from the small number of galaxies in the associated group and it has had the wings of the distribution trimmed (see \S5 in Paper 3). The number of galaxies in the overdensity is a strong function of the depth and completeness of the spectroscopic coverage, which varies across the sky. Nevertheless, it is reassuring that in a gross sense the two measurements track each other (Figure \ref{fig:environment}).

We use these two measures in concert to provide a broad guideline regarding the local environment of each UDG candidate for which a distance is estimated using this technique. Because of the significant scatter present in Figure \ref{fig:environment}, we opt to set joint limits and limit our environmental designation to solely rich ($\sigma_v >$ 500 km s$^{-1}$ and $N \ge 15$) or poor ($\sigma_v \le 500$ km s$^{-1}$ and $N < 15)$ environments.

\subsection{Estimating Masses}
\label{sec:masses}

We follow the procedure described by \cite{zb} and applied by \cite{zgc} to estimate total masses for as many of our UDG candidates with estimated distances as possible. The procedure divides into two steps. 

First, we use a galaxy scaling relation to estimate the velocity dispersion of each galaxy. 
The scaling relation connecting $r_e$, the surface brightness within $r_e$, $I_e$, and velocity dispersion, $\sigma_v$ across all types of stellar systems has been discussed extensively in a set of papers \citep{zfm1,zfm2}. Having all but one of these parameters, the last can be estimated using the relation. 

Second, we use the velocity dispersion to estimate the mass within $r_e$, subtract the contribution of baryons within $r_e$, and find the NFW \citep{nfw} model that has the corresponding dark matter mass within $r_e$. This estimate we test using globular cluster abundances for six UDGs. Both of these steps are described in more detail next.

\begin{figure}[ht]
\begin{center}
\includegraphics[width=0.47\textwidth]{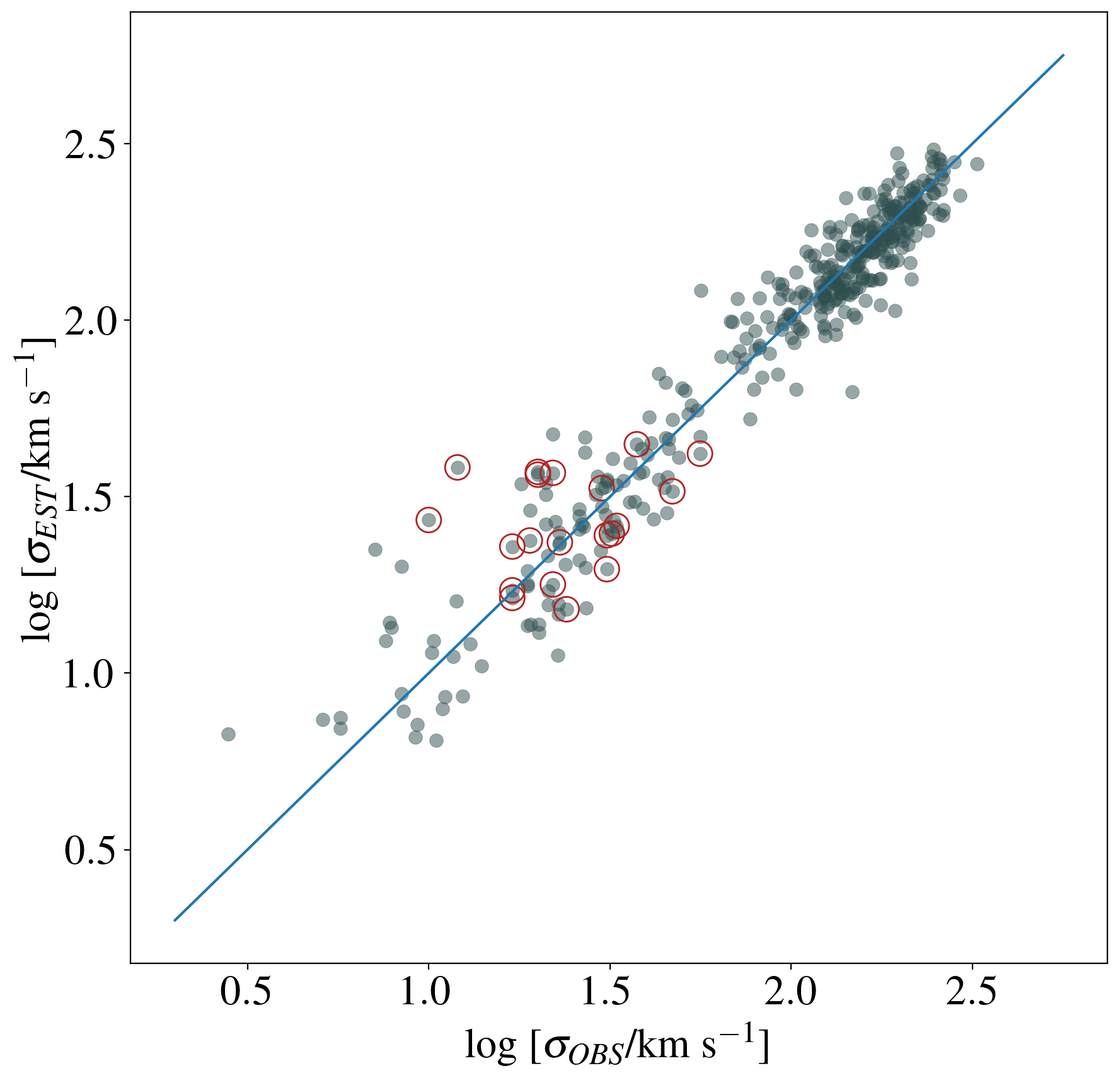}
\end{center}
\vskip -.5cm
\caption{Comparison of estimated velocity dispersions of dispersion supported galaxies of various masses to the spectroscopically measured values \citep{jorgensen,geha,chilingarian08,mieske,collins}. Twenty UDGs with velocity dispersion measurements from the literature \citep{Beasley+2016a,vdk17,toloba,chilingarian19,martin-navarro,vdk19,forbes21,gannon} are highlighted with larger, open red circles. The dispersion about the 1:1 line for UDGs corresponds to a velocity dispersion scatter of 11 km s$^{-1}$.} 
\label{fig:fm}
\end{figure}

\subsubsection{Velocity Dispersion}

In Figure \ref{fig:fm} we compare our estimated velocity dispersions, $\sigma_{EST}$, against  spectroscopically measured values for a wide range of galaxies from the literature (references in Figure caption), and highlight UDGs (also drawn from published data). We use the empirically-determined coefficients for the scaling relation given in \cite{zb}. With larger spectroscopic data sets these coefficients should be further refined, particularly to better constrain the relationship in the region of parameter space populated by  UDGs.

The UDGs follow the general trend on the 1:1 line and do not exhibit larger scatter than other galaxies of similar velocity dispersion. The rms scatter for the UDGs corresponds to 11 km s$^{-1}$, roughly a $\sqrt{2}$ increase over the typical quoted uncertainty in the spectroscopic measurements, suggesting the estimated values have comparable precision to the spectroscopic ones. We propose that 11 km s$^{-1}$ is a rough upper limit on the 1$\sigma$ uncertainty of our velocity dispersion estimates.  Of course, estimates have the potential for catastrophic errors and should always be treated with caution. We will return to this issue later, but the two UDGs with the lowest observed velocity dispersions suggest that the empirical relationship may be breaking down for low velocity dispersion UDGs. We proceed to calcute $\sigma_{EST}$ for every UDG candidate in the catalog with an estimated redshift. These estimates are included in the catalog but should be treated with caution and await further confirmation.

\subsubsection{Modeling Total Masses}

We estimate total masses by adopting an underlying NFW potential, determining the total mass within $r_e$ using the Wolf mass estimator \citep{wolf}, subtracting the stellar mass using a color dependent stellar mass-to-light \citep{roediger} within $r_e$\footnote{See \cite{du} for an extensive discussion of mass-to-light ratios for low surface brightness galaxies. Although there are a variety of uncertainties, they do not identify any gross difference between the mass-to-light ratio vs. color relations for low and high surface brightness galaxies. Any possible differences are also mitigated by our study of galaxies where the stellar contribution to the mass within $r_e$ is subdominant.}, and finding the NFW profile that best matches the residual mass within $r_e$. This approach assumes that all of the baryons within $r_e$ are in stars and that there is no adiabatic contraction of the halo due to the baryons. To help satisfy the latter assumption we apply the method only to systems where the dark matter mass fraction within $r_e$ is at least 50\%. This effectively limits the technique to dwarf galaxies. Nevertheless, even if the original underlying potential was NFW-like, baryonic effects other than contraction, such as feedback, could also affect the shape of the potential \citep{sawala}. Even so, there is yet no compelling argument that NFW potentials are inappropriate for UDGs \citep{Sales+2020}, but whether this is an accurate assumption remains an open question.

There are limited ways to test the results of the method, including the assumption of the NFW profile, in galaxies in general --- and even fewer in UDGs. We will use the relationship between the number of globular clusters, $N_{\rm GC}$, (or the total mass of the globular cluster population, which for a universal luminosity function is directly related to the number) and total galaxy mass, M$_h$, that is now well established for the general galaxy population \citep{blakeslee, spitler, georgiev, harris13, hudson, forbes16, harris17, forbes18, burkert, zgc}. The relationship has already been assumed to hold for UDGs and used to estimate the total mass of many individual UDGs \citep[e.g.,][]{Beasley+2016b,pl,amorisco}.

\begin{figure}[ht]
\begin{center}
\includegraphics[width=0.47\textwidth]{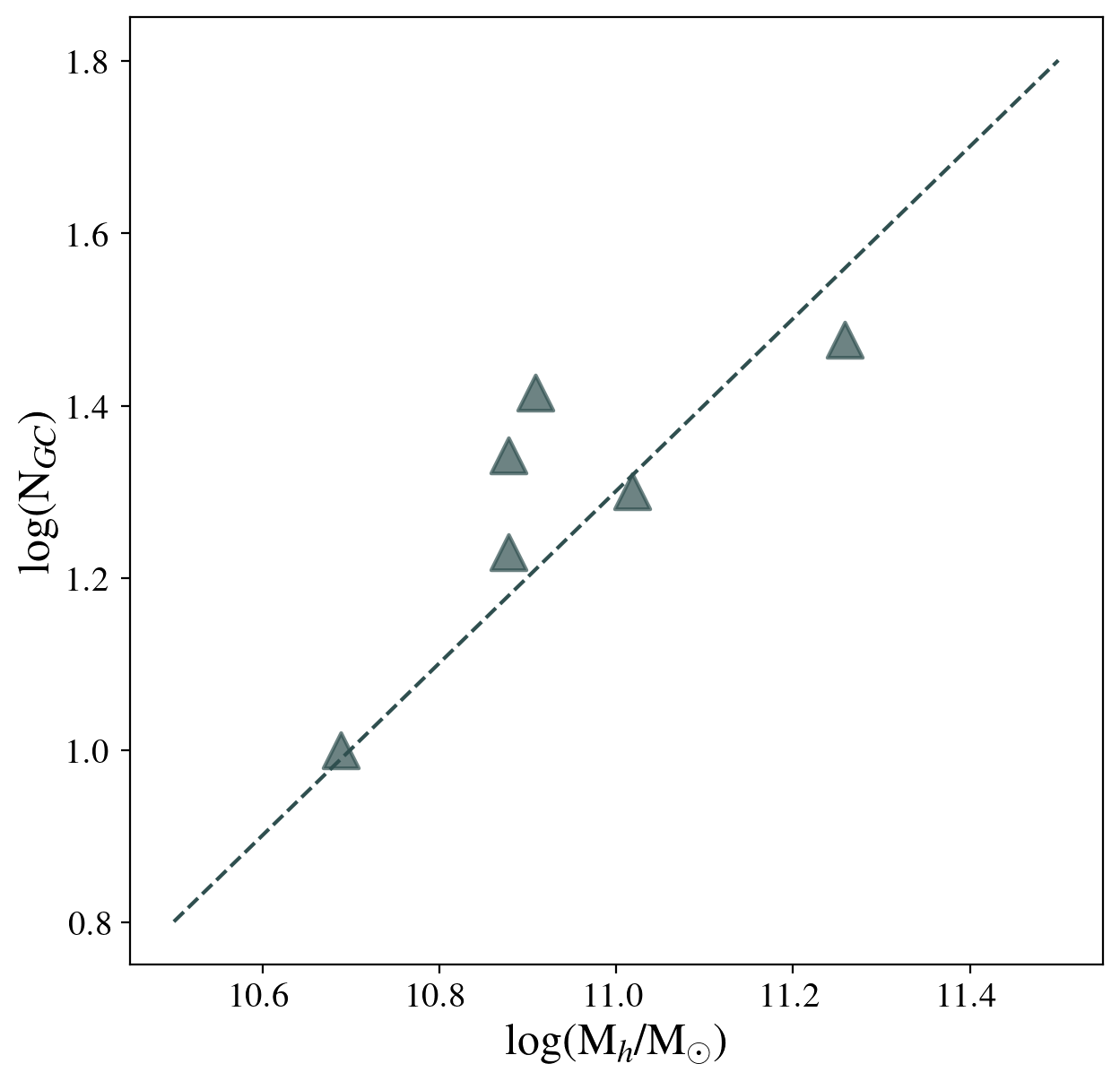}
\end{center}
\vskip -.5cm
\caption{The relationship between the number of globular clusters and total mass. The number of globular clusters come from \cite{Saifollahi+2021}  and the masses from our estimates. The dashed line is not a fit but rather the published relation from \cite{burkert}. The comparison indicates a modest offset and acceptable precision for our estimated masses (mean offset from the relation is 15\% and the scatter corresponds to an uncertainty in the mass of 25\%). However, as detailed in the text, some exceptions to this behavior are known from other studies.}
\label{fig:gc}
\end{figure}

In Figure \ref{fig:gc} we compare N$_{\rm GC}$ for a sample of six well-characterized UDGs \citep{Saifollahi+2021} and our estimates of  $M_h$. The line plotted in the Figure is the relation obtained from the general galaxy population, $M_h = (5\times10^9$M$_\odot$)N$_{GC}$ \citep{burkert}. If the published relation is taken as being correct, then our M$_h$ estimates are underestimates of the total mass by $\sim$15\% on average and have a scatter of $\sim$25\%. If this evaluation of the accuracy and precision of the method is even close to being correct, it suggests that the method provides an exciting way forward to estimate masses for large numbers of low mass galaxies, including UDGs. We present in the catalog our estimates for M$_h$ for the subset of 1436 UDG candidates where we can proceed with the calculation.

We chose to focus the N$_{GC}$-M$_h$ comparison on the \cite{Saifollahi+2021} sample because N$_{GC}$ is based on deep HST imaging and a uniform treatment of selection and completeness corrections across the sample. However, a similar comparison has catastrophic failures when one considers a broader set of measurements. First, the two UDGs associated with  NGC 1052 that have little or no apparent dark matter within $r_e$ \citep{vdkdm1,vdkdm2} are clear outliers in this relation, independent of the mass estimation approach \citep{vdk-gc}. They have substantial globular cluster populations but a low total mass. Although a variety of formation scenarios have been proposed for these systems, some are sufficiently fine-tuned \citep[e.g.,][]{vdk-bullet} to suggest that such objects should be exceedingly rare across the entire SMUDGes sample. 

Also concerning is the case of NGC 5846 UDG1 \citep{forbes19}, although any one object may not be representative. 
Based on ground-based imaging, the galaxy appears to have a bountiful GC population \citep[17 clusters initially identified, completeness corrections would more than double that number;][]{forbes21}, with 11 now spectroscopically confirmed \citep{muller2020}. Using the scaling relation and the structural parameters presented by \cite{forbes19}, we obtain $\sigma_{EST} = 23$ km s$^{-1}$, in agreement with the published spectroscopic measurement, $\sigma_v = 17\pm$2 km s$^{-1}$ \citep{forbes21}. However, our subsequent estimate of the total mass is only $1.7\times10^{10}$ M$_\odot$ in comparison to their estimate, based on N$_{GC}$, of $\sim 2\times 10^{11}$ M$_\odot$. This is a factor of ten discrepant, suggesting  N$_{GC}$ would need to be $\sim 10$ times smaller for us to place the galaxy on the relationship in Figure \ref{fig:gc}.
There seems to be little room for observational error at this level.
Subsequent {\sl HST} observations \citep{muller2021,danieli2022} confirm the large number of globular clusters and adopting a subsequent measurement of  $\sigma_v$ using globular clusters \citep[$\sigma_v = 9.4^{+7.0}_{-5.4}$ km s$^{-1}$;][]{muller2020} would make the discrepancy even worse because it lowers the inferred dynamical mass. 

 An alternative interpretation was suggested by \cite{forbes21} because they too found a conflict between the mass within $r_e$ and the total mass. They suggest that the dark matter potential may be cored. If this is true, we would be underestimating the total mass by fitting an NFW model. Determining whether such discrepancies are common or isolated may be a way to learn about the dark matter potentials of these systems. We cannot make further progress here, but more careful study of the population of GC-rich low surface brightness galaxies is essential.

\section{Star Formation Efficiencies}
\label{sec:discussion}

We now address  whether the integrated star formation efficiency of UDGs is different than that of galaxies of similar total masses. We move forward with some trepidation given the chain of steps required to estimate masses, but hopeful that in the mean the masses are sufficiently accurate. Confirming this assertion is a priority for future work but we also discuss below why we do not anticipate qualitative changes in the results.

\begin{figure}[ht]
\begin{center}
\includegraphics[width=0.47\textwidth]{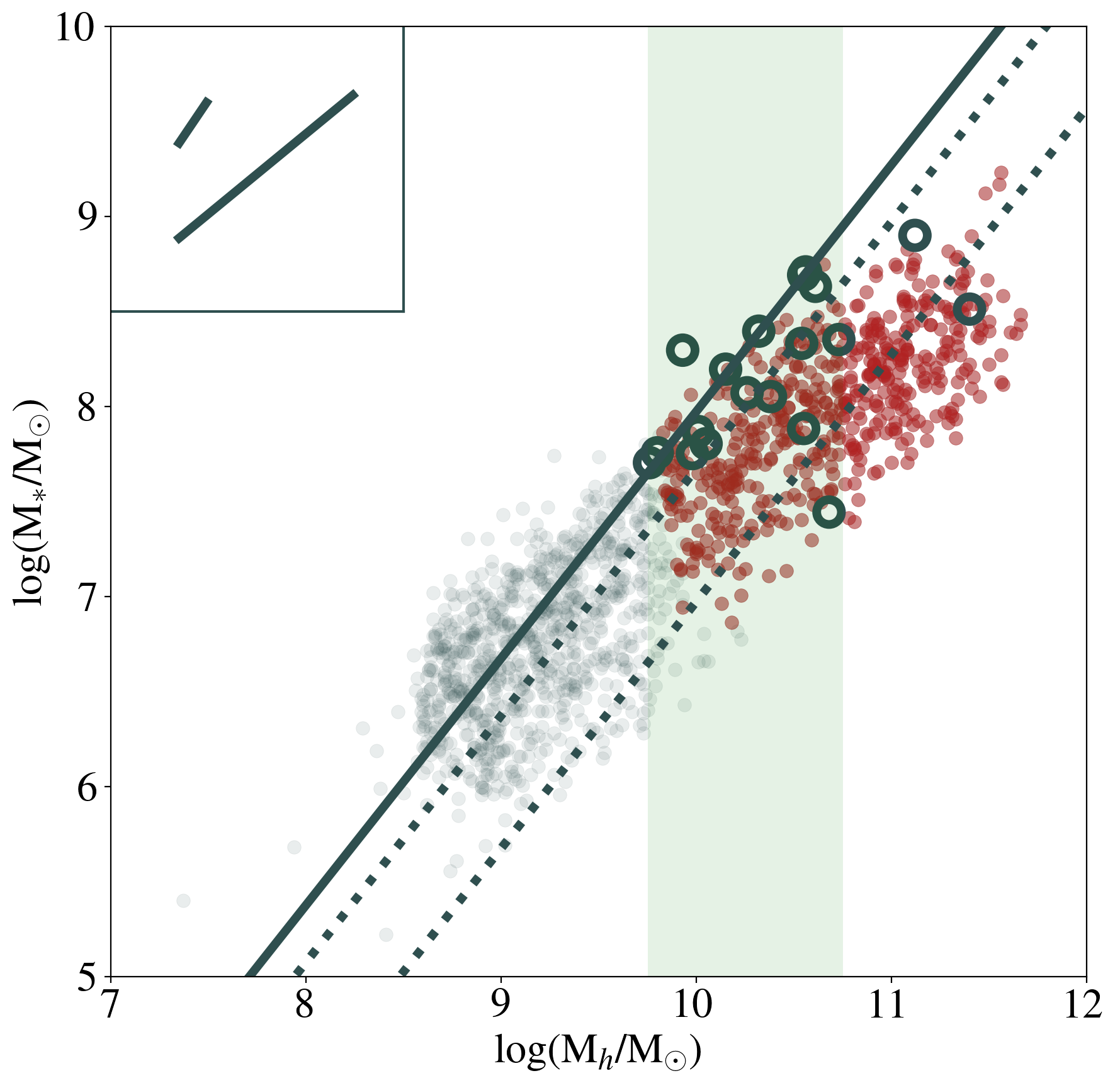}
\end{center}
\vskip -.5cm
\caption{The SMHM relation for SMUDGes sources. The red, filled symbols represent UDGs ($r_e \ge 1.5$ kpc), which corresponds closely to a cut in M$_h$. The lighter colored points represent the SMUDGes that do not meet the UDG size criterion. The solid line represents the mean relation for a wide set of dwarf galaxies drawn from the literature \citep{zb}. The dotted lines represent a decrease in the stellar mass by a factor of 2 and 10. The large unfilled dark circles represent the results for a sample of UDGs with spectroscopically-measured velocity dispersions and independent photometry. The shaded region highlights the range of M$_h$ where we compare the samples in the text. The lines in the inset represent how an individual galaxy will move with a 10\% and 100\% error in the distance. Small errors move points along the published relation but significantly overestimating the distance moves points in the direction populated by the largest UDGs. The UDGs are relatively inefficient integrated star forming galaxies at a given M$_h$. At a quantitative level, the results await further validation of the methodology for estimating $\sigma_v$ and M$_h$.}
\label{fig:smhm}
\end{figure}

In Figure \ref{fig:smhm} we compare the stellar mass-halo mass (SMHM) relation for UDGs and other smaller low surface brightness galaxies in SMUDGes to the mean relation found using the same approach for the general dwarf galaxy population \citep{zb}. The UDGs are offset from the mean relation in the sense that they are star-deficient at a given M$_h$. 
The magnitude of the effect is largest for those with the largest size or M$_h$. As we consider galaxies with lower values of M$_h$ the SMUDGes sources eventually merge onto the mean relation. On average, the offset for UDGs with $9.73 < $log(M$_h/$M$_\odot) < 10.7$ (the shaded region in Figure \ref{fig:smhm}) is 0.61$\pm$0.02 dex (a factor of $\sim$ 4), but for the most massive UDGs the integrated star formation efficiency is roughly an order of magnitude lower than for the general galaxy population. Over the full mass range of UDGs we find an average factor of 7 deficiency. The range of M$_h$ we find for UDGs, $10^{10} <$ M$_h$/M$_\odot < 10^{11.5}$, is mostly in line with certain theoretical expectations \citep[10$^{10}-10^{11}$ M$_\odot$;][]{DiCintio+2017,rong} and consistent with limits from gravitational lensing \citep{sifon}, providing some further encouragement for our mass estimates.

In the same Figure we also plot the sample of UDGs that we used in Figure \ref{fig:z_est} to test the velocity dispersion estimates. Here we use the spectroscopic velocity dispersion measurements and the independent values of $r_e$, distance, and magnitudes provided by the source references and derive M$_h$ as we do for our sample. Despite the fact that this is a much smaller sample and mean trends are more difficult to quantify, these points also preferentially fall below the fiducial SMHM relation. The offset within the shaded region in Figure \ref{fig:smhm} is not quite as large as for the SMUDGes sample in the same mass range (mean offset  is $0.22\pm0.10)$ dex (a factor of $\sim$ 1.7 deficient in stars in comparison to a factor of 4). We defined the shaded region to maximize the overlap between our sample and literature sample in M$_h$ and provide a fair baseline for comparison. For the entire literature sample we find an average offset of 
$0.29\pm0.10$ dex.

An alternate interpretation, if one posits that galaxies have the same mean stellar to dark matter mass ratio, is that the dark matter profiles differ systematically between 'normal' galaxies and UDGs in this mass range, such that we are incorrectly estimating the masses for at least one of these two types of galaxies. However, the sense of the difference is that the UDGs would have to have a more concentrated dark matter profile than the normal galaxies, a trend which seems at odds with their larger sizes. We prefer the interpretation that the star formation efficiencies differ.

Although the results from our sample and the literature sample both suggest a deficit of stars in UDGs, the quantitative results disagree, suggesting the presence of systematic uncertainties.
On our side of the equation, we face potential systematics in the estimation of the distances and velocity dispersions.
We expect that 30\% of the sample has incorrect distances (\S\ref{sec:distances}) and know that distance errors tend to scatter sources at an angle to the fiducial SMHM relation that is consistent with that seen in Figure \ref{fig:smhm}  \citep{zb}. Furthermore,  systematic differences in $r_e$ measurements among samples can lead to offset differences.  For example, if we reduce our measurements of $r_e$ by 30\% we decrease the measured offset to our fiducial SMHM to $0.41\pm0.02$ dex, which is now roughly only 2$\sigma$ discrepant with the literature value. On the literature side of the equation, observational biases (spectroscopy is most likely to succeed for UDGs with higher surface brightnesses or for star-forming UDGs exhibiting emission lines) could lead to unrepresentative samples that favor systems with higher relative stellar masses. Given the uncertainties in this discussion, we broadly claim that the stellar deficiency lies somewhere between a factor of 1.5 to 4 in the mass range of the comparison and 1.5 to 7 for the full UDG mass range. 

There are various possible causes for the relative low integrated star formation efficiency in UDGs within the existing set of formation scenarios. For example, gas stripping would naturally lead to less star formation over the lifetime of the galaxy and a high specific angular momentum could halt some gas from collapsing and reaching the densities required for star formation. The one set of models that can be excluded by this result are those that explain UDGs solely by redistributing the stars
to larger radii at late times. However, most models that invoke dynamical interactions as a key component of dwarf galaxy evolution suggest that a redistribution of stars occurs in conjunction with the loss of stars from the system \citep[e.g.,][]{penarrubia,tomozeiu,Carleton+2019}. 

The most extreme offsets are at the largest masses. 
The estimates of M$_h$ can surely be incorrect, but appear unlikely to be overestimated by a factor of 10. The agreement between the estimated and measured velocity dispersions shown in Figure \ref{fig:fm} suggests that there is no large overall bias in the dispersion estimates, eliminating this specific issue as a concern. As mentioned previously, on further examination of \ref{fig:fm} there is some concern about a possible systematic failure of our dispersion estimates for log $\sigma_v < 1.25$ but this is not the regime of the most massive UDGs and the dispersion estimates agree well with measurements above log $\sigma_v = 1.25$. The previous discussion regarding N$_{GC}$ and the extrapolated total masses suggests similar concurrence with independent estimates (Figure \ref{fig:gc}). 

Another concern is the effect of distance errors, which propagate into both M$_*$ and M$_h$. We show in Figure \ref{fig:smhm} the direction and amplitude of shifts due to 10\% and 100\% distance errors. For small errors  the points slide basically along the direction of the published relation and do not result in offsets from it. However, as the errors grow the shift pushes the data away from the published relation in a manner consistent with what we find if many of our distances are gross overestimates. Of course, 100\% errors fall into the category of catastrophic distance errors and we only expect such errors for 30\% the sample. If we remove the 30\% of UDGs with the largest masses, the largest remaining UDG still has M$_h > 10^{11}$ M$_\odot$. For this mass, the offset from the fiducial SMHM relation is still about an order  of magnitude. However, such errors may explain why there is a difference in the mean offset between the full sample and that composed of  galaxies with spectroscopic distances and $\sigma_v$. 

We could also be overestimating masses if the adopted potential profile is incorrect. However, by selecting a cuspy profile, NFW, we are likely to underestimate rather than overestimate the mass, as may indeed be the case for NGC 5846 UDG1. There are ways to contaminate the mass estimates, such as with the presence of central massive black hole, but that contamination would have to be systemic and different in a relative sense to what is occurring in galaxies of similar stellar mass. Citing Occam's razor, we conclude that UDGs are the relative star-deficient tail of the galaxy distribution at the corresponding values of M$_h$. This does not exclude the possibility that they are also the physically large tail of the population and that those two aspects are physically related.

\begin{figure}[ht]
\begin{center}
\includegraphics[width=0.47\textwidth]{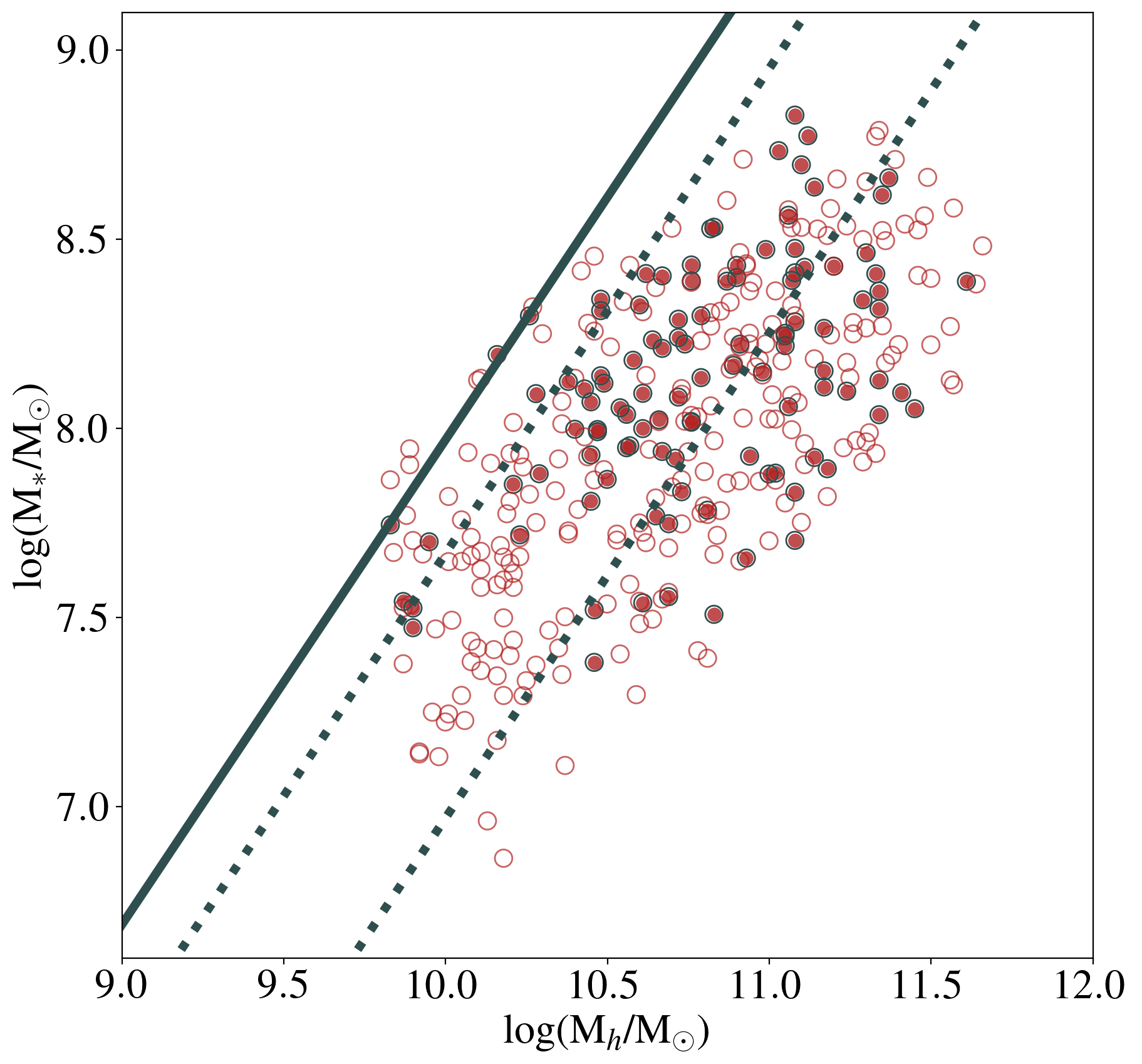}
\end{center}
\vskip -.5cm
\caption{The stellar mass-halo mass relation for UDGs in poor and rich environments. The solid circles represent those UDGs in rich environments and open circles those in poor ones, with environment defined as described in the text. The solid line represents the mean relation for the general population from \cite{zb}.
The dotted lines represent a decrease in the stellar mass by a factor of 2 and 10. The uncertainties due to distance errors are as shown in Figure 10.}
\label{fig:smhm-environment}
\end{figure}

As a test of the effect of interactions and the role of local environment, we present the SMHM relation of UDGs separately for those in rich and poor environments as defined in \S\ref{sec:environment} in Figure \ref{fig:smhm-environment}. There is no clear distinction between the two populations. Of course absence of evidence is not evidence of absence. For example, the outer portions of the dark matter halo could have been stripped away in the subset of UDGs in high density environments and any gas reservoir may have been tidally or ram pressure stripped as well. Neither of these events would necessarily show up in our comparison. What the agreement between the two SMHM relations does demonstrate is that environment has not affected the structure within $r_e$ sufficiently to differentially affect our estimates of $\sigma_v$ and M$_h$. As such, we interpret the agreement to mean that dynamical processes are unlikely to affect $r_e$ and hence to be directly related to the creation of UDGs. This conclusion is independently supported by the near linearity of the N$_{UDG}$ vs. 
halo mass relation from 10$^{12}$ to 10$^{15}$ M$_\odot$ \citep[e.g.,][]{Karunakaran+2023}, although  we again stress that given the likely heterogeneous nature of the UDG populations, models that produce a variety of UDGs \citep[e.g.,][]{Sales+2020} may have enough flexibility to match these disparate observations. Our results at least present empirical benchmarks against which to test such models.

\begin{figure*}[ht]
\begin{center}
\includegraphics[width=0.95\textwidth]{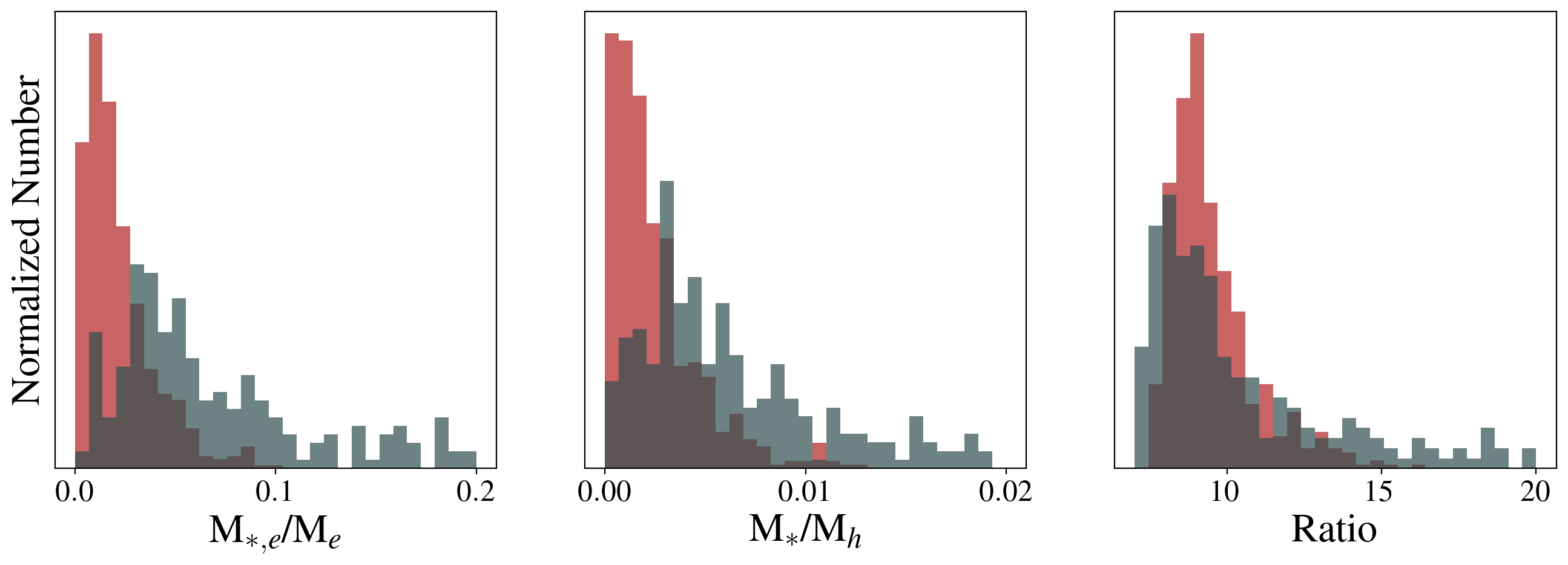}
\end{center}
\vskip -.5cm
\caption{Comparison of UDGs (red) and the general galaxy population (grey) with $10 < $M$_h/M_\odot < 11.5$. The first panel shows the stellar mass to total mass ratio within $r_e$, while the second shows the same ratio for the halo radius, $r_h$. Finally, the third panel shows the distribution of the ratio of these two quantities. Smaller ratios indicate lower concentration of stars relative to dark matter.} 
\label{fig:concentration}
\end{figure*}

To assess whether UDGs are both star-deficient and ``puffier", we compare the ratio of stellar to total mass within $r_e$, M$_{*,e}/$M$_e$, and within the virial radius, M$_*/$M$_h$,
for UDGs and the general galaxy population over a limited range of M$_h$ ($10 <$ M$_h/$M$_\odot < 11.5$). If UDGs are simply deficient in stars but have no structural differences, then these ratios should change by the same factor within $r_e$ and $r_{200}$. We present that comparison in Figure \ref{fig:concentration}.

The distributions of M$_{*,e}/$M$_e$ and M$_{*}/$M$_h$ are quite similar and that impression is confirmed in the third panel, which plots the ratio of the value at $r_e$ to that at $r_h$ for the two galaxy populations. The UDG population does not exhibit significantly different behavior, suggesting that size differences in the stellar component are not a primary factor. In fact, the ratio peaks at slightly smaller values for the general population, probably due to the larger relative contribution of the stellar mass within $r_e$ in the general galaxy population. We conclude that whatever processes are involved in forming a UDG they lead primarily to a decrease in the integrated star formation rate in galaxies of comparable total mass.

\section{Summary}

In this last of the SMUDGes catalog papers we present the full UDG candidate catalog drawn from the Legacy Survey DR9 release \citep{Dey+2019} following our described procedure and selection. The catalog contains 7,070 candidates with $\mu_{0,g} \ge 24$ mag arcsec$^{-2}$ and $r_e \ge 5.3$ arcsec as measured using single S\'ersic models and our particular masking algorithms. 
After visual examination, we consider 265 of these to be poor candidates and flag them as such. Using the parameters of each candidate, we estimate and tabulate the completeness fraction across the full survey for similar galaxies using artificial source simulations. The catalog is highly incomplete for sources with $r_e \gtrsim 20$ arcsec due to our choice of filtering. The mean overall completeness for galaxies like those already in the catalog is close to 50\%. 

We estimate and tabulate distances for a subsample of 1,525 candidates using the distance-by-association technique described in Paper III, associating  candidates projected onto the Coma, Fornax, and Virgo clusters with those clusters, and using the \cite{Kadowaki+21} compilation of spectroscopic redshifts. Because of the overrepresentation of Virgo and Fornax galaxies in this list, the fraction of UDGs ($r_e \ge 1.5$ kpc) in this list is small (40\%) resulting in a total of 585 UDGs, but they are distributed across the sky. 
They have total magnitudes typically in the range $-17 \lesssim$ M$_r$ $ \lesssim -14$ and are typically red, $g-r \sim 0.6$, although there is a blue population that extends to $g-r \sim 0.1$.

We estimate and tabulate total masses, M$_h$, using an approach presented by \cite{zb}. Although the approach is speculative for UDGs and needs further validation, it provides a way forward at the current time. It provides accurate (to within 25\%) mass estimates in comparison to those from the number of globular clusters for a sample of 6 UDGs \citep{Saifollahi+2021}. We find that UDGs have total masses in the range of $10^{10} \lesssim $ M$_h/$M$_\odot \lesssim 10^{11.5}$.
Comparing the stellar mass-halo mass (SMHM) relation of UDGs to that of the general population over the same range of M$_h$ from \cite{zb}, we find that UDGs are increasingly star-deficient with increasing M$_h$. The average deficit of stars varies as a function of galaxy size and total mass and likely lies somewhere between a factor 1.5 and 7 for the UDG sample as a whole.

We conclude that whatever processes are involved in forming UDGs, they do not simply reorganize the stars to larger radii but instead result in a measurable decrease, up to an order of magnitude, in the integrated star formation rates relative to other galaxies of the same total mass ($10^{10} < $M$_h$/M$_\odot$ $< 10^{11.5}$). This deficiency does not have a detectable environmental dependence.

%\appendix
%\section{Modifications to Simulation processing}
%\label{app:Simchanges}
%To obtain adequate statistics from our simulations, we create the dataset for the models by taking median values drawn from those passing automated classification (input dataset) using randomly placed, fixed-length windows for each of the four parameters, thus creating a four dimensional parameter space.  We define our window sizes as the integer fraction of the input data set and, while the window length may be quite large for a single parameter, it provides a much smaller number of data points in the 4D parameter space.  Although larger windows allow more robust statistics, they limit the parameter range resulting in some of the science candidates falling outside of the parameter space defined by the model.  Using the criteria described in Paper II, we find that the more uniform depth distribution in MzLS/BASS relative to DECaLS allows us to use smaller window sizes of one ninth of the input data set rather than the previous one seventh and still provides models that fit the data at least as well while expanding the parameter space.

\begin{acknowledgments}

DZ, RD, JK, DJK, and HZ acknowledge financial support from NSF AST-1713841 and AST-2006785. DZ thanks the Astronomy Department at Columbia University for hosting him during his sabbatical. AD's research is supported by NOIRLab. KS acknowledges funding from the Natural Sciences and Engineering Research Council of Canada (NSERC). AK acknowledges financial support from the grant CEX2021-001131-S funded by MCIN/AEI/ 10.13039/501100011033 and from the grant POSTDOC\_21\_00845 funded by the Economic Transformation, Industry, Knowledge and Universities Council of the Regional Government of Andalusia. An allocation of computer time from the UA Research Computing High Performance Computing (HPC) at the University of Arizona and the prompt assistance of the associated computer support group is gratefully acknowledged.

This research has made use of the NASA/IPAC Extragalactic Database (NED), which is operated by the Jet Propulsion Laboratory, California Institute of Technology, under contract with NASA.

The Legacy Surveys consist of three individual and complementary projects: the Dark Energy Camera Legacy Survey (DECaLS; Proposal ID \#2014B-0404; PIs: David Schlegel and Arjun Dey), the Beijing-Arizona Sky Survey (BASS; NOAO Prop. ID \#2015A-0801; PIs: Zhou Xu and Xiaohui Fan), and the Mayall z-band Legacy Survey (MzLS; Prop. ID \#2016A-0453; PI: Arjun Dey). DECaLS, BASS and MzLS together include data obtained, respectively, at the Blanco telescope, Cerro Tololo Inter-American Observatory, NSF's NOIRLab; the Bok telescope, Steward Observatory, University of Arizona; and the Mayall telescope, Kitt Peak National Observatory, NOIRLab. Pipeline processing and analyses of the data were supported by NOIRLab and the Lawrence Berkeley National Laboratory (LBNL). The Legacy Surveys project is honored to be permitted to conduct astronomical research on Iolkam Du'ag (Kitt Peak), a mountain with particular significance to the Tohono O'odham Nation.

NOIRLab is operated by the Association of Universities for Research in Astronomy (AURA) under a cooperative agreement with the National Science Foundation. LBNL is managed by the Regents of the University of California under contract to the U.S. Department of Energy.

This project used data obtained with the Dark Energy Camera (DECam), which was constructed by the Dark Energy Survey (DES) collaboration. Funding for the DES Projects has been provided by the U.S. Department of Energy, the U.S. National Science Foundation, the Ministry of Science and Education of Spain, the Science and Technology Facilities Council of the United Kingdom, the Higher Education Funding Council for England, the National Center for Supercomputing Applications at the University of Illinois at Urbana-Champaign, the Kavli Institute of Cosmological Physics at the University of Chicago, Center for Cosmology and Astro-Particle Physics at the Ohio State University, the Mitchell Institute for Fundamental Physics and Astronomy at Texas A\&M University, Financiadora de Estudos e Projetos, Fundacao Carlos Chagas Filho de Amparo, Financiadora de Estudos e Projetos, Fundacao Carlos Chagas Filho de Amparo a Pesquisa do Estado do Rio de Janeiro, Conselho Nacional de Desenvolvimento Cientifico e Tecnologico and the Ministerio da Ciencia, Tecnologia e Inovacao, the Deutsche Forschungsgemeinschaft and the Collaborating Institutions in the Dark Energy Survey. The Collaborating Institutions are Argonne National Laboratory, the University of California at Santa Cruz, the University of Cambridge, Centro de Investigaciones Energeticas, Medioambientales y Tecnologicas-Madrid, the University of Chicago, University College London, the DES-Brazil Consortium, the University of Edinburgh, the Eidgenossische Technische Hochschule (ETH) Zurich, Fermi National Accelerator Laboratory, the University of Illinois at Urbana-Champaign, the Institut de Ciencies de l'Espai (IEEC/CSIC), the Institut de Fisica d’Altes Energies, Lawrence Berkeley National Laboratory, the Ludwig Maximilians Universitat Munchen and the associated Excellence Cluster Universe, the University of Michigan, NSF’s NOIRLab, the University of Nottingham, the Ohio State University, the University of Pennsylvania, the University of Portsmouth, SLAC National Accelerator Laboratory, Stanford University, the University of Sussex, and Texas A\&M University.

BASS is a key project of the Telescope Access Program (TAP), which has been funded by the National Astronomical Observatories of China, the Chinese Academy of Sciences (the Strategic Priority Research Program “The Emergence of Cosmological Structures” Grant \#XDB09000000), and the Special Fund for Astronomy from the Ministry of Finance. The BASS is also supported by the External Cooperation Program of Chinese Academy of Sciences (Grant \#4A11KYSB20160057), and Chinese National Natural Science Foundation (Grant \#12120101003, \#11433005).

The Legacy Survey team makes use of data products from the Near-Earth Object Wide-field Infrared Survey Explorer (NEOWISE), which is a project of the Jet Propulsion Laboratory/California Institute of Technology. NEOWISE is funded by the National Aeronautics and Space Administration.

The Legacy Surveys imaging of the DESI footprint is supported by the Director, Office of Science, Office of High Energy Physics of the U.S. Department of Energy under Contract No. DE-AC02-05CH1123, by the National Energy Research Scientific Computing Center, a DOE Office of Science User Facility under the same contract; and by the U.S. National Science Foundation, Division of Astronomical Sciences under Contract No. AST-0950945 to NOAO.

\end{acknowledgments}

\medskip
\facilities{Blanco, Mayall, Bok}

\software{
Astropy              \citep{astropy1, astropy2},
astroquery           \citep{astroquery},
GALFIT               \citep{peng},
keras                \citep{keras},
lmfit                \citep{newville},
Matplotlib           \citep{matplotlib},
NumPy                \citep{numpy},
pandas               \citep{pandas},
sep                  \citep{sep},
Source Extractor     \citep{bertin},
SciPy                \citep{scipy1, scipy2},
SWarp                \citep{Swarp}}

%\bibliography{references.bib}{}
\bibliography{refs.bib}

\begin{thebibliography}{}
\expandafter\ifx\csname natexlab\endcsname\relax\def\natexlab#1{#1}\fi
\providecommand{\url}[1]{\href{#1}{#1}}
\providecommand{\dodoi}[1]{doi:~\href{http://doi.org/#1}{\nolinkurl{#1}}}
\providecommand{\doeprint}[1]{\href{http://ascl.net/#1}{\nolinkurl{http://ascl.net/#1}}}
\providecommand{\doarXiv}[1]{\href{https://arxiv.org/abs/#1}{\nolinkurl{https://arxiv.org/abs/#1}}}

\bibitem[{{Amorisco} \& {Loeb}(2016)}]{Amorisco+2016}
{Amorisco}, N.~C., \& {Loeb}, A. 2016, \mnras, 459, L51,
  \dodoi{10.1093/mnrasl/slw055}

\bibitem[{{Amorisco} {et~al.}(2018){Amorisco}, {Monachesi}, {Agnello}, \&
  {White}}]{amorisco}
{Amorisco}, N.~C., {Monachesi}, A., {Agnello}, A., \& {White}, S.~D.~M. 2018,
  \mnras, 475, 4235, \dodoi{10.1093/mnras/sty116}

\bibitem[{{Astropy Collaboration} {et~al.}(2013){Astropy Collaboration},
  {Robitaille}, {Tollerud}, {Greenfield}, {Droettboom}, {Bray}, {Aldcroft},
  {Davis}, {Ginsburg}, {Price-Whelan}, {Kerzendorf}, {Conley}, {Crighton},
  {Barbary}, {Muna}, {Ferguson}, {Grollier}, {Parikh}, {Nair}, {Unther},
  {Deil}, {Woillez}, {Conseil}, {Kramer}, {Turner}, {Singer}, {Fox}, {Weaver},
  {Zabalza}, {Edwards}, {Azalee Bostroem}, {Burke}, {Casey}, {Crawford},
  {Dencheva}, {Ely}, {Jenness}, {Labrie}, {Lim}, {Pierfederici}, {Pontzen},
  {Ptak}, {Refsdal}, {Servillat}, \& {Streicher}}]{astropy1}
{Astropy Collaboration}, {Robitaille}, T.~P., {Tollerud}, E.~J., {et~al.} 2013,
  \aap, 558, A33, \dodoi{10.1051/0004-6361/201322068}

\bibitem[{{Astropy Collaboration} {et~al.}(2018){Astropy Collaboration},
  {Price-Whelan}, {Sip{\H{o}}cz}, {G{\"u}nther}, {Lim}, {Crawford}, {Conseil},
  {Shupe}, {Craig}, {Dencheva}, {Ginsburg}, {VanderPlas}, {Bradley},
  {P{\'e}rez-Su{\'a}rez}, {de Val-Borro}, {Aldcroft}, {Cruz}, {Robitaille},
  {Tollerud}, {Ardelean}, {Babej}, {Bach}, {Bachetti}, {Bakanov}, {Bamford},
  {Barentsen}, {Barmby}, {Baumbach}, {Berry}, {Biscani}, {Boquien}, {Bostroem},
  {Bouma}, {Brammer}, {Bray}, {Breytenbach}, {Buddelmeijer}, {Burke},
  {Calderone}, {Cano Rodr{\'\i}guez}, {Cara}, {Cardoso}, {Cheedella}, {Copin},
  {Corrales}, {Crichton}, {D'Avella}, {Deil}, {Depagne}, {Dietrich}, {Donath},
  {Droettboom}, {Earl}, {Erben}, {Fabbro}, {Ferreira}, {Finethy}, {Fox},
  {Garrison}, {Gibbons}, {Goldstein}, {Gommers}, {Greco}, {Greenfield},
  {Groener}, {Grollier}, {Hagen}, {Hirst}, {Homeier}, {Horton}, {Hosseinzadeh},
  {Hu}, {Hunkeler}, {Ivezi{\'c}}, {Jain}, {Jenness}, {Kanarek}, {Kendrew},
  {Kern}, {Kerzendorf}, {Khvalko}, {King}, {Kirkby}, {Kulkarni}, {Kumar},
  {Lee}, {Lenz}, {Littlefair}, {Ma}, {Macleod}, {Mastropietro}, {McCully},
  {Montagnac}, {Morris}, {Mueller}, {Mumford}, {Muna}, {Murphy}, {Nelson},
  {Nguyen}, {Ninan}, {N{\"o}the}, {Ogaz}, {Oh}, {Parejko}, {Parley}, {Pascual},
  {Patil}, {Patil}, {Plunkett}, {Prochaska}, {Rastogi}, {Reddy Janga},
  {Sabater}, {Sakurikar}, {Seifert}, {Sherbert}, {Sherwood-Taylor}, {Shih},
  {Sick}, {Silbiger}, {Singanamalla}, {Singer}, {Sladen}, {Sooley},
  {Sornarajah}, {Streicher}, {Teuben}, {Thomas}, {Tremblay}, {Turner},
  {Terr{\'o}n}, {van Kerkwijk}, {de la Vega}, {Watkins}, {Weaver}, {Whitmore},
  {Woillez}, {Zabalza}, \& {Astropy Contributors}}]{astropy2}
{Astropy Collaboration}, {Price-Whelan}, A.~M., {Sip{\H{o}}cz}, B.~M., {et~al.}
  2018, \aj, 156, 123, \dodoi{10.3847/1538-3881/aabc4f}

\bibitem[{Barbary(2016)}]{sep}
Barbary, K. 2016, {SEP: Source Extractor as a library},
  \dodoi{10.21105/joss.00058}

\bibitem[{{Beasley} {et~al.}(2016){Beasley}, {Romanowsky}, {Pota}, {Navarro},
  {Martinez Delgado}, {Neyer}, \& {Deich}}]{Beasley+2016a}
{Beasley}, M.~A., {Romanowsky}, A.~J., {Pota}, V., {et~al.} 2016, \apjl, 819,
  L20, \dodoi{10.3847/2041-8205/819/2/L20}

\bibitem[{{Beasley} \& {Trujillo}(2016)}]{Beasley+2016b}
{Beasley}, M.~A., \& {Trujillo}, I. 2016, \apj, 830, 23,
  \dodoi{10.3847/0004-637X/830/1/23}

\bibitem[{{Bertin} \& {Arnouts}(1996)}]{bertin}
{Bertin}, E., \& {Arnouts}, S. 1996, \aaps, 117, 393,
  \dodoi{10.1051/aas:1996164}

\bibitem[{{Bertin} {et~al.}(2002){Bertin}, {Mellier}, {Radovich}, {Missonnier},
  {Didelon}, \& {Morin}}]{Swarp}
{Bertin}, E., {Mellier}, Y., {Radovich}, M., {et~al.} 2002, Astronomical
  Society of the Pacific Conference Series, Vol. 281, {The TERAPIX Pipeline},
  ed. D.~A. {Bohlender}, D.~{Durand}, \& T.~H. {Handley}, 228

\bibitem[{{Blakeslee} {et~al.}(1997){Blakeslee}, {Tonry}, \&
  {Metzger}}]{blakeslee}
{Blakeslee}, J.~P., {Tonry}, J.~L., \& {Metzger}, M.~R. 1997, \aj, 114, 482,
  \dodoi{10.1086/118488}

\bibitem[{{Burkert} \& {Forbes}(2020)}]{burkert}
{Burkert}, A., \& {Forbes}, D.~A. 2020, \aj, 159, 56,
  \dodoi{10.3847/1538-3881/ab5b0e}

\bibitem[{Carleton {et~al.}(2019)Carleton, Errani, Cooper, Kaplinghat,
  Peñarrubia, \& Guo}]{Carleton+2019}
Carleton, T., Errani, R., Cooper, M., {et~al.} 2019, Monthly Notices of the
  Royal Astronomical Society, 485, 382, \dodoi{10.1093/mnras/stz383}

\bibitem[{{Chan} {et~al.}(2018){Chan}, {Kere{\v{s}}}, {Wetzel}, {Hopkins},
  {Faucher-Gigu{\`e}re}, {El-Badry}, {Garrison-Kimmel}, \&
  {Boylan-Kolchin}}]{Chan+2018}
{Chan}, T.~K., {Kere{\v{s}}}, D., {Wetzel}, A., {et~al.} 2018, \mnras, 478,
  906, \dodoi{10.1093/mnras/sty1153}

\bibitem[{{Chilingarian} {et~al.}(2019){Chilingarian}, {Afanasiev}, {Grishin},
  {Fabricant}, \& {Moran}}]{chilingarian19}
{Chilingarian}, I.~V., {Afanasiev}, A.~V., {Grishin}, K.~A., {Fabricant}, D.,
  \& {Moran}, S. 2019, \apj, 884, 79, \dodoi{10.3847/1538-4357/ab4205}

\bibitem[{{Chilingarian} {et~al.}(2008){Chilingarian}, {Cayatte}, {Durret},
  {Adami}, {Balkowski}, {Chemin}, {Lagan{\'a}}, \& {Prugniel}}]{chilingarian08}
{Chilingarian}, I.~V., {Cayatte}, V., {Durret}, F., {et~al.} 2008, \aap, 486,
  85, \dodoi{10.1051/0004-6361:20078709}

\bibitem[{{Chollet, ~F. and Keras Team}(2015)}]{keras}
{Chollet, ~F. and Keras Team}. 2015, Keras

\bibitem[{{Collins} {et~al.}(2014){Collins}, {Chapman}, {Rich}, {Ibata},
  {Martin}, {Irwin}, {Bate}, {Lewis}, {Pe{\~n}arrubia}, {Arimoto}, {Casey},
  {Ferguson}, {Koch}, {McConnachie}, \& {Tanvir}}]{collins}
{Collins}, M. L.~M., {Chapman}, S.~C., {Rich}, R.~M., {et~al.} 2014, \apj, 783,
  7, \dodoi{10.1088/0004-637X/783/1/7}

\bibitem[{{Danieli} {et~al.}(2022){Danieli}, {van Dokkum}, {Trujillo-Gomez},
  {Kruijssen}, {Romanowsky}, {Carlsten}, {Shen}, {Li}, {Abraham}, {Brodie},
  {Conroy}, {Gannon}, \& {Greco}}]{danieli2022}
{Danieli}, S., {van Dokkum}, P., {Trujillo-Gomez}, S., {et~al.} 2022, \apjl,
  927, L28, \dodoi{10.3847/2041-8213/ac590a}

\bibitem[{Dey {et~al.}(2016)Dey, Rabinowitz, Karcher, Bebek, Baltay,
  Sprayberry, Valdes, Stupak, Donaldson, Emmet, Hurteau, Abareshi, Marshall,
  Lang, Fitzpatrick, Daly, Joyce, Schlegel, Schweiker, Allen, Blum, \&
  Levi}]{mosaic}
Dey, A., Rabinowitz, D., Karcher, A., {et~al.} 2016, in Ground-based and
  Airborne Instrumentation for Astronomy VI, ed. C.~J. Evans, L.~Simard, \&
  H.~Takami, Vol. 9908, International Society for Optics and Photonics (SPIE),
  750 -- 757, \dodoi{10.1117/12.2231488}

\bibitem[{{Dey} {et~al.}(2019){Dey}, {Schlegel}, {Lang}, {Blum}, {Burleigh},
  {Fan}, {Findlay}, {Finkbeiner}, {Herrera}, {Juneau}, {Landriau}, {Levi},
  {McGreer}, {Meisner}, {Myers}, {Moustakas}, {Nugent}, {Patej}, {Schlafly},
  {Walker}, {Valdes}, {Weaver}, {Y{\`e}che}, {Zou}, {Zhou}, {Abareshi},
  {Abbott}, {Abolfathi}, {Aguilera}, {Alam}, {Allen}, {Alvarez}, {Annis},
  {Ansarinejad}, {Aubert}, {Beechert}, {Bell}, {BenZvi}, {Beutler}, {Bielby},
  {Bolton}, {Brice{\~n}o}, {Buckley-Geer}, {Butler}, {Calamida}, {Carlberg},
  {Carter}, {Casas}, {Castander}, {Choi}, {Comparat}, {Cukanovaite}, {Delubac},
  {DeVries}, {Dey}, {Dhungana}, {Dickinson}, {Ding}, {Donaldson}, {Duan},
  {Duckworth}, {Eftekharzadeh}, {Eisenstein}, {Etourneau}, {Fagrelius},
  {Farihi}, {Fitzpatrick}, {Font-Ribera}, {Fulmer}, {G{\"a}nsicke},
  {Gaztanaga}, {George}, {Gerdes}, {Gontcho}, {Gorgoni}, {Green}, {Guy},
  {Harmer}, {Hernandez}, {Honscheid}, {Huang}, {James}, {Jannuzi}, {Jiang},
  {Joyce}, {Karcher}, {Karkar}, {Kehoe}, {Kneib}, {Kueter-Young}, {Lan},
  {Lauer}, {Le Guillou}, {Le Van Suu}, {Lee}, {Lesser}, {Perreault Levasseur},
  {Li}, {Mann}, {Marshall}, {Mart{\'\i}nez-V{\'a}zquez}, {Martini}, {du Mas des
  Bourboux}, {McManus}, {Meier}, {M{\'e}nard}, {Metcalfe},
  {Mu{\~n}oz-Guti{\'e}rrez}, {Najita}, {Napier}, {Narayan}, {Newman}, {Nie},
  {Nord}, {Norman}, {Olsen}, {Paat}, {Palanque-Delabrouille}, {Peng},
  {Poppett}, {Poremba}, {Prakash}, {Rabinowitz}, {Raichoor}, {Rezaie},
  {Robertson}, {Roe}, {Ross}, {Ross}, {Rudnick}, {Safonova}, {Saha},
  {S{\'a}nchez}, {Savary}, {Schweiker}, {Scott}, {Seo}, {Shan}, {Silva},
  {Slepian}, {Soto}, {Sprayberry}, {Staten}, {Stillman}, {Stupak}, {Summers},
  {Sien Tie}, {Tirado}, {Vargas-Maga{\~n}a}, {Vivas}, {Wechsler}, {Williams},
  {Yang}, {Yang}, {Yapici}, {Zaritsky}, {Zenteno}, {Zhang}, {Zhang}, {Zhou}, \&
  {Zhou}}]{Dey+2019}
{Dey}, A., {Schlegel}, D.~J., {Lang}, D., {et~al.} 2019, \aj, 157, 168,
  \dodoi{10.3847/1538-3881/ab089d}

\bibitem[{{Di Cintio} {et~al.}(2017){Di Cintio}, {Brook}, {Dutton},
  {Macci{\`o}}, {Obreja}, \& {Dekel}}]{DiCintio+2017}
{Di Cintio}, A., {Brook}, C.~B., {Dutton}, A.~A., {et~al.} 2017, \mnras, 466,
  L1, \dodoi{10.1093/mnrasl/slw210}

\bibitem[{{Du} {et~al.}(2020){Du}, {Cheng}, {Zheng}, \& {Wu}}]{du}
{Du}, W., {Cheng}, C., {Zheng}, Z., \& {Wu}, H. 2020, \aj, 159, 138,
  \dodoi{10.3847/1538-3881/ab6efb}

\bibitem[{{Erkal} {et~al.}(2019){Erkal}, {Belokurov}, {Laporte}, {Koposov},
  {Li}, {Grillmair}, {Kallivayalil}, {Price-Whelan}, {Evans}, {Hawkins},
  {Hendel}, {Mateu}, {Navarro}, {del Pino}, {Slater}, {Sohn}, \& {Orphan Aspen
  Treasury Collaboration}}]{erkal}
{Erkal}, D., {Belokurov}, V., {Laporte}, C.~F.~P., {et~al.} 2019, \mnras, 487,
  2685, \dodoi{10.1093/mnras/stz1371}

\bibitem[{{Forbes} {et~al.}(2016){Forbes}, {Alabi}, {Romanowsky}, {Brodie},
  {Strader}, {Usher}, \& {Pota}}]{forbes16}
{Forbes}, D.~A., {Alabi}, A., {Romanowsky}, A.~J., {et~al.} 2016, \mnras, 458,
  L44, \dodoi{10.1093/mnrasl/slw015}

\bibitem[{{Forbes} {et~al.}(2019){Forbes}, {Gannon}, {Couch}, {Iodice},
  {Spavone}, {Cantiello}, {Napolitano}, \& {Schipani}}]{forbes19}
{Forbes}, D.~A., {Gannon}, J., {Couch}, W.~J., {et~al.} 2019, \aap, 626, A66,
  \dodoi{10.1051/0004-6361/201935499}

\bibitem[{{Forbes} {et~al.}(2021){Forbes}, {Gannon}, {Romanowsky}, {Alabi},
  {Brodie}, {Couch}, \& {Ferr{\'e}-Mateu}}]{forbes21}
{Forbes}, D.~A., {Gannon}, J.~S., {Romanowsky}, A.~J., {et~al.} 2021, \mnras,
  500, 1279, \dodoi{10.1093/mnras/staa3289}

\bibitem[{{Forbes} {et~al.}(2018){Forbes}, {Read}, {Gieles}, \&
  {Collins}}]{forbes18}
{Forbes}, D.~A., {Read}, J.~I., {Gieles}, M., \& {Collins}, M. L.~M. 2018,
  \mnras, 481, 5592, \dodoi{10.1093/mnras/sty2584}

\bibitem[{{Gannon} {et~al.}(2022){Gannon}, {Forbes}, {Romanowsky},
  {Ferr{\'e}-Mateu}, {Couch}, {Brodie}, {Huang}, {Janssens}, \&
  {Okabe}}]{gannon}
{Gannon}, J.~S., {Forbes}, D.~A., {Romanowsky}, A.~J., {et~al.} 2022, \mnras,
  510, 946, \dodoi{10.1093/mnras/stab3297}

\bibitem[{{Geha} {et~al.}(2003){Geha}, {Guhathakurta}, \& {van der
  Marel}}]{geha}
{Geha}, M., {Guhathakurta}, P., \& {van der Marel}, R.~P. 2003, \aj, 126, 1794,
  \dodoi{10.1086/377624}

\bibitem[{{Georgiev} {et~al.}(2010){Georgiev}, {Puzia}, {Goudfrooij}, \&
  {Hilker}}]{georgiev}
{Georgiev}, I.~Y., {Puzia}, T.~H., {Goudfrooij}, P., \& {Hilker}, M. 2010,
  \mnras, 406, 1967, \dodoi{10.1111/j.1365-2966.2010.16802.x}

\bibitem[{{Ginsburg} {et~al.}(2019){Ginsburg}, {Sip{\H{o}}cz}, {Brasseur},
  {Cowperthwaite}, {Craig}, {Deil}, {Guillochon}, {Guzman}, {Liedtke}, {Lian
  Lim}, {Lockhart}, {Mommert}, {Morris}, {Norman}, {Parikh}, {Persson},
  {Robitaille}, {Segovia}, {Singer}, {Tollerud}, {de Val-Borro}, {Valtchanov},
  {Woillez}, {Astroquery Collaboration}, \& {a subset of astropy
  Collaboration}}]{astroquery}
{Ginsburg}, A., {Sip{\H{o}}cz}, B.~M., {Brasseur}, C.~E., {et~al.} 2019, \aj,
  157, 98, \dodoi{10.3847/1538-3881/aafc33}

\bibitem[{{Goto} {et~al.}(2023){Goto}, {Zaritsky}, {Karunakaran},
  {Donnerstein}, \& {Sand}}]{goto}
{Goto}, H., {Zaritsky}, D., {Karunakaran}, A., {Donnerstein}, R., \& {Sand},
  David, J. 2023, \apj

\bibitem[{{Greco} {et~al.}(2018){Greco}, {Goulding}, {Greene}, {Strauss},
  {Huang}, {Kim}, \& {Komiyama}}]{Greco+2018b}
{Greco}, J.~P., {Goulding}, A.~D., {Greene}, J.~E., {et~al.} 2018, \apj, 866,
  112, \dodoi{10.3847/1538-4357/aae0f4}

\bibitem[{{Green}(2018)}]{green}
{Green}, G. 2018, The Journal of Open Source Software, 3, 695,
  \dodoi{10.21105/joss.00695}

\bibitem[{{Greene} {et~al.}(2022){Greene}, {Greco}, {Goulding}, {Huang},
  {Kado-Fong}, {Danieli}, {Li}, {Kim}, {Komiyama}, {Leauthaud}, {MacArthur}, \&
  {Sif{\'o}n}}]{greene}
{Greene}, J.~E., {Greco}, J.~P., {Goulding}, A.~D., {et~al.} 2022, \apj, 933,
  150, \dodoi{10.3847/1538-4357/ac7238}

\bibitem[{{Grishin} {et~al.}(2021){Grishin}, {Chilingarian}, {Afanasiev},
  {Fabricant}, {Katkov}, {Moran}, \& {Yagi}}]{grishin}
{Grishin}, K.~A., {Chilingarian}, I.~V., {Afanasiev}, A.~V., {et~al.} 2021,
  Nature Astronomy, 5, 1308, \dodoi{10.1038/s41550-021-01470-5}

\bibitem[{{Harris} {et~al.}(2017){Harris}, {Blakeslee}, \& {Harris}}]{harris17}
{Harris}, W.~E., {Blakeslee}, J.~P., \& {Harris}, G. L.~H. 2017, \apj, 836, 67,
  \dodoi{10.3847/1538-4357/836/1/67}

\bibitem[{{Harris} {et~al.}(2013){Harris}, {Harris}, \& {Alessi}}]{harris13}
{Harris}, W.~E., {Harris}, G. L.~H., \& {Alessi}, M. 2013, \apj, 772, 82,
  \dodoi{10.1088/0004-637X/772/2/82}

\bibitem[{{Hinshaw} {et~al.}(2013){Hinshaw}, {Larson}, {Komatsu}, {Spergel},
  {Bennett}, {Dunkley}, {Nolta}, {Halpern}, {Hill}, {Odegard}, {Page}, {Smith},
  {Weiland}, {Gold}, {Jarosik}, {Kogut}, {Limon}, {Meyer}, {Tucker}, {Wollack},
  \& {Wright}}]{hinshaw}
{Hinshaw}, G., {Larson}, D., {Komatsu}, E., {et~al.} 2013, \apjs, 208, 19,
  \dodoi{10.1088/0067-0049/208/2/19}

\bibitem[{{Hudson} {et~al.}(2014){Hudson}, {Harris}, \& {Harris}}]{hudson}
{Hudson}, M.~J., {Harris}, G.~L., \& {Harris}, W.~E. 2014, \apjl, 787, L5,
  \dodoi{10.1088/2041-8205/787/1/L5}

\bibitem[{{Hunter}(2007)}]{matplotlib}
{Hunter}, J.~D. 2007, Computing in Science and Engineering, 9, 90,
  \dodoi{10.1109/MCSE.2007.55}

\bibitem[{Jones {et~al.}(2001)Jones, Oliphant, P., \& et~al.}]{jones}
Jones, E., Oliphant, T., P., P., \& et~al. 2001, {SciPy}: Open source

\bibitem[{{Jorgensen} {et~al.}(1996){Jorgensen}, {Franx}, \&
  {Kjaergaard}}]{jorgensen}
{Jorgensen}, I., {Franx}, M., \& {Kjaergaard}, P. 1996, \mnras, 280, 167,
  \dodoi{10.1093/mnras/280.1.167}

\bibitem[{{Kadowaki} {et~al.}(2021){Kadowaki}, {Zaritsky}, {Donnerstein}, {RS},
  {Karunakaran}, \& {Spekkens}}]{Kadowaki+21}
{Kadowaki}, J., {Zaritsky}, D., {Donnerstein}, R.~L., {et~al.} 2021, \apj, 923,
  257, \dodoi{10.3847/1538-4357/ac2948}

\bibitem[{{Karunakaran} \& {Zaritsky}(2023)}]{Karunakaran+2023}
{Karunakaran}, A., \& {Zaritsky}, D. 2023, \mnras, 519, 884,
  \dodoi{10.1093/mnras/stac3622}

\bibitem[{{Koda} {et~al.}(2015){Koda}, {Yagi}, {Yamanoi}, \&
  {Komiyama}}]{Koda+2015}
{Koda}, J., {Yagi}, M., {Yamanoi}, H., \& {Komiyama}, Y. 2015, \apjl, 807, L2,
  \dodoi{10.1088/2041-8205/807/1/L2}

\bibitem[{{Leisman} {et~al.}(2017){Leisman}, {Haynes}, {Janowiecki},
  {Hallenbeck}, {J{\'o}zsa}, {Giovanelli}, {Adams}, {Bernal Neira}, {Cannon},
  {Janesh}, {Rhode}, \& {Salzer}}]{Leisman+2017}
{Leisman}, L., {Haynes}, M.~P., {Janowiecki}, S., {et~al.} 2017, \apj, 842,
  133, \dodoi{10.3847/1538-4357/aa7575}

\bibitem[{{Li} {et~al.}(2022){Li}, {Greene}, {Greco}, {Huang}, {Melchior},
  {Beaton}, {Casey}, {Danieli}, {Goulding}, {Joseph}, {Kado-Fong}, {Kim}, \&
  {MacArthur}}]{li22}
{Li}, J., {Greene}, J.~E., {Greco}, J.~P., {et~al.} 2022, arXiv e-prints,
  arXiv:2210.14994, \dodoi{10.48550/arXiv.2210.14994}

\bibitem[{{Martin} {et~al.}(2019){Martin}, {Kaviraj}, {Laigle}, {Devriendt},
  {Jackson}, {Peirani}, {Dubois}, {Pichon}, \& {Slyz}}]{Martin+2019}
{Martin}, G., {Kaviraj}, S., {Laigle}, C., {et~al.} 2019, \mnras, 485, 796,
  \dodoi{10.1093/mnras/stz356}

\bibitem[{{Mart{\'\i}n-Navarro} {et~al.}(2019){Mart{\'\i}n-Navarro},
  {Romanowsky}, {Brodie}, {Ferr{\'e}-Mateu}, {Alabi}, {Forbes}, {Sharina},
  {Villaume}, {Pandya}, \& {Martinez-Delgado}}]{martin-navarro}
{Mart{\'\i}n-Navarro}, I., {Romanowsky}, A.~J., {Brodie}, J.~P., {et~al.} 2019,
  \mnras, 484, 3425, \dodoi{10.1093/mnras/stz252}

\bibitem[{{Mart{\'\i}nez-Delgado} {et~al.}(2016){Mart{\'\i}nez-Delgado},
  {L{\"a}sker}, {Sharina}, {Toloba}, {Fliri}, {Beaton}, {Valls-Gabaud},
  {Karachentsev}, {Chonis}, {Grebel}, {Forbes}, {Romanowsky},
  {Gallego-Laborda}, {Teuwen}, {G{\'o}mez-Flechoso}, {Wang}, {Guhathakurta},
  {Kaisin}, \& {Ho}}]{Martinez-Delgado+2016}
{Mart{\'\i}nez-Delgado}, D., {L{\"a}sker}, R., {Sharina}, M., {et~al.} 2016,
  \aj, 151, 96, \dodoi{10.3847/0004-6256/151/4/96}

\bibitem[{{McKinney}(2010)}]{pandas}
{McKinney}, W. 2010, Proceedings of the 9th Python in Science Conference, 51

\bibitem[{{Meisner} \& {Finkbeiner}(2014)}]{meisner}
{Meisner}, A.~M., \& {Finkbeiner}, D.~P. 2014, \apj, 781, 5,
  \dodoi{10.1088/0004-637X/781/1/5}

\bibitem[{{Mieske} {et~al.}(2008){Mieske}, {Hilker}, {Jord{\'a}n}, {Infante},
  {Kissler-Patig}, {Rejkuba}, {Richtler}, {C{\^o}t{\'e}}, {Baumgardt}, {West},
  {Ferrarese}, \& {Peng}}]{mieske}
{Mieske}, S., {Hilker}, M., {Jord{\'a}n}, A., {et~al.} 2008, \aap, 487, 921,
  \dodoi{10.1051/0004-6361:200810077}

\bibitem[{{Millman} \& {Aivazis}(2011)}]{scipy2}
{Millman}, K.~J., \& {Aivazis}, M. 2011, Computing in Science and Engineering,
  13, 9, \dodoi{10.1109/MCSE.2011.36}

\bibitem[{{M{\"u}ller} {et~al.}(2020){M{\"u}ller}, {Marleau}, {Duc}, {Habas},
  {Fensch}, {Emsellem}, {Poulain}, {Lim}, {Agnello}, {Durrell}, {Paudel},
  {S{\'a}nchez-Janssen}, \& {van der Burg}}]{muller2020}
{M{\"u}ller}, O., {Marleau}, F.~R., {Duc}, P.-A., {et~al.} 2020, \aap, 640,
  A106, \dodoi{10.1051/0004-6361/202038351}

\bibitem[{{M{\"u}ller} {et~al.}(2021){M{\"u}ller}, {Durrell}, {Marleau}, {Duc},
  {Lim}, {Posti}, {Agnello}, {S{\'a}nchez-Janssen}, {Poulain}, {Habas},
  {Emsellem}, {Paudel}, {van der Burg}, \& {Fensch}}]{muller2021}
{M{\"u}ller}, O., {Durrell}, P.~R., {Marleau}, F.~R., {et~al.} 2021, \apj, 923,
  9, \dodoi{10.3847/1538-4357/ac2831}

\bibitem[{{Navarro} {et~al.}(1997){Navarro}, {Frenk}, \& {White}}]{nfw}
{Navarro}, J.~F., {Frenk}, C.~S., \& {White}, S. D.~M. 1997, \apj, 490, 493,
  \dodoi{10.1086/304888}

\bibitem[{Newville {et~al.}(2014)Newville, Stensitzki, Allen, \&
  Ingargiola}]{newville}
Newville, M., Stensitzki, T., Allen, D.~B., \& Ingargiola, A. 2014, {LMFIT:
  Non-Linear Least-Square Minimization and Curve-Fitting for Python}

\bibitem[{{Oke}(1964)}]{oke1}
{Oke}, J.~B. 1964, \apj, 140, 689, \dodoi{10.1086/147960}

\bibitem[{{Oke} \& {Gunn}(1983)}]{oke2}
{Oke}, J.~B., \& {Gunn}, J.~E. 1983, \apj, 266, 713, \dodoi{10.1086/160817}

\bibitem[{{Oliphant}(2007)}]{scipy1}
{Oliphant}, T.~E. 2007, Computing in Science and Engineering, 9, 10,
  \dodoi{10.1109/MCSE.2007.58}

\bibitem[{{Pe{\~n}arrubia} {et~al.}(2008){Pe{\~n}arrubia}, {Navarro}, \&
  {McConnachie}}]{penarrubia}
{Pe{\~n}arrubia}, J., {Navarro}, J.~F., \& {McConnachie}, A.~W. 2008, \apj,
  673, 226, \dodoi{10.1086/523686}

\bibitem[{Pedregosa {et~al.}(2011)Pedregosa, Varoquaux, Gramfort, Michel,
  Thirion, Grisel, Blondel, Prettenhofer, Weiss, Dubourg, Vanderplas, Passos,
  Cournapeau, Brucher, Perrot, \& Duchesnay}]{sklearn}
Pedregosa, F., Varoquaux, G., Gramfort, A., {et~al.} 2011, Journal of Machine
  Learning Research, 12, 2825

\bibitem[{{Peng} {et~al.}(2002){Peng}, {Ho}, {Impey}, \& {Rix}}]{peng}
{Peng}, C.~Y., {Ho}, L.~C., {Impey}, C.~D., \& {Rix}, H.-W. 2002, \aj, 124,
  266, \dodoi{10.1086/340952}

\bibitem[{{Peng} \& {Lim}(2016)}]{pl}
{Peng}, E.~W., \& {Lim}, S. 2016, \apjl, 822, L31,
  \dodoi{10.3847/2041-8205/822/2/L31}

\bibitem[{{Planck Collaboration} {et~al.}(2014){Planck Collaboration},
  {Abergel}, {Ade}, {Aghanim}, {Alves}, {Aniano}, {Armitage-Caplan}, {Arnaud},
  {Ashdown}, {Atrio-Barand ela}, {Aumont}, {Baccigalupi}, {Banday}, {Barreiro},
  {Bartlett}, {Battaner}, {Benabed}, {Beno{\^\i}t}, {Benoit-L{\'e}vy},
  {Bernard}, {Bersanelli}, {Bielewicz}, {Bobin}, {Bock}, {Bonaldi}, {Bond},
  {Borrill}, {Bouchet}, {Boulanger}, {Bridges}, {Bucher}, {Burigana}, {Butler},
  {Cardoso}, {Catalano}, {Chamballu}, {Chary}, {Chiang}, {Chiang},
  {Christensen}, {Church}, {Clemens}, {Clements}, {Colombi}, {Colombo},
  {Combet}, {Couchot}, {Coulais}, {Crill}, {Curto}, {Cuttaia}, {Danese},
  {Davies}, {Davis}, {de Bernardis}, {de Rosa}, {de Zotti}, {Delabrouille},
  {Delouis}, {D{\'e}sert}, {Dickinson}, {Diego}, {Dole}, {Donzelli},
  {Dor{\'e}}, {Douspis}, {Draine}, {Dupac}, {Efstathiou}, {En{\ss}lin},
  {Eriksen}, {Falgarone}, {Finelli}, {Forni}, {Frailis}, {Fraisse},
  {Franceschi}, {Galeotta}, {Ganga}, {Ghosh}, {Giard}, {Giardino},
  {Giraud-H{\'e}raud}, {Gonz{\'a}lez-Nuevo}, {G{\'o}rski}, {Gratton},
  {Gregorio}, {Grenier}, {Gruppuso}, {Guillet}, {Hansen}, {Hanson}, {Harrison},
  {Helou}, {Henrot-Versill{\'e}}, {Hern{\'a}ndez-Monteagudo}, {Herranz},
  {Hildebrand t}, {Hivon}, {Hobson}, {Holmes}, {Hornstrup}, {Hovest},
  {Huffenberger}, {Jaffe}, {Jaffe}, {Jewell}, {Joncas}, {Jones}, {Juvela},
  {Keih{\"a}nen}, {Keskitalo}, {Kisner}, {Knoche}, {Knox}, {Kunz},
  {Kurki-Suonio}, {Lagache}, {L{\"a}hteenm{\"a}ki}, {Lamarre}, {Lasenby},
  {Laureijs}, {Lawrence}, {Leonardi}, {Le{\'o}n-Tavares}, {Lesgourgues},
  {Levrier}, {Liguori}, {Lilje}, {Linden-V{\o}rnle}, {L{\'o}pez-Caniego},
  {Lubin}, {Mac{\'\i}as-P{\'e}rez}, {Maffei}, {Maino}, {Mand olesi}, {Maris},
  {Marshall}, {Martin}, {Mart{\'\i}nez-Gonz{\'a}lez}, {Masi}, {Massardi},
  {Matarrese}, {Matthai}, {Mazzotta}, {McGehee}, {Melchiorri}, {Mendes},
  {Mennella}, {Migliaccio}, {Mitra}, {Miville-Desch{\^e}nes}, {Moneti},
  {Montier}, {Morgante}, {Mortlock}, {Munshi}, {Murphy}, {Naselsky}, {Nati},
  {Natoli}, {Netterfield}, {N{\o}rgaard-Nielsen}, {Noviello}, {Novikov},
  {Novikov}, {Osborne}, {Oxborrow}, {Paci}, {Pagano}, {Pajot}, {Paladini},
  {Paoletti}, {Pasian}, {Patanchon}, {Perdereau}, {Perotto}, {Perrotta},
  {Piacentini}, {Piat}, {Pierpaoli}, {Pietrobon}, {Plaszczynski},
  {Pointecouteau}, {Polenta}, {Ponthieu}, {Popa}, {Poutanen}, {Pratt},
  {Pr{\'e}zeau}, {Prunet}, {Puget}, {Rachen}, {Reach}, {Rebolo}, {Reinecke},
  {Remazeilles}, {Renault}, {Ricciardi}, {Riller}, {Ristorcelli}, {Rocha},
  {Rosset}, {Roudier}, {Rowan-Robinson}, {Rubi{\~n}o-Mart{\'\i}n}, {Rusholme},
  {Sandri}, {Santos}, {Savini}, {Scott}, {Seiffert}, {Shellard}, {Spencer},
  {Starck}, {Stolyarov}, {Stompor}, {Sudiwala}, {Sunyaev}, {Sureau}, {Sutton},
  {Suur-Uski}, {Sygnet}, {Tauber}, {Tavagnacco}, {Terenzi}, {Toffolatti},
  {Tomasi}, {Tristram}, {Tucci}, {Tuovinen}, {T{\"u}rler}, {Umana},
  {Valenziano}, {Valiviita}, {Van Tent}, {Verstraete}, {Vielva}, {Villa},
  {Vittorio}, {Wade}, {Wandelt}, {Welikala}, {Ysard}, {Yvon}, {Zacchei}, \&
  {Zonca}}]{planck}
{Planck Collaboration}, {Abergel}, A., {Ade}, P.~A.~R., {et~al.} 2014, \aap,
  571, A11, \dodoi{10.1051/0004-6361/201323195}

\bibitem[{{Prole} {et~al.}(2019){Prole}, {van der Burg}, {Hilker}, \&
  {Davies}}]{Prole+2019}
{Prole}, D.~J., {van der Burg}, R.~F.~J., {Hilker}, M., \& {Davies}, J.~I.
  2019, \mnras, 488, 2143, \dodoi{10.1093/mnras/stz1843}

\bibitem[{{Roediger} \& {Courteau}(2015)}]{roediger}
{Roediger}, J.~C., \& {Courteau}, S. 2015, \mnras, 452, 3209,
  \dodoi{10.1093/mnras/stv1499}

\bibitem[{{Rom{\'a}n} \& {Trujillo}(2017)}]{Roman+2017}
{Rom{\'a}n}, J., \& {Trujillo}, I. 2017, \mnras, 468, 4039,
  \dodoi{10.1093/mnras/stx694}

\bibitem[{{Rong} {et~al.}(2017){Rong}, {Guo}, {Gao}, {Liao}, {Xie}, {Puzia},
  {Sun}, \& {Pan}}]{rong}
{Rong}, Y., {Guo}, Q., {Gao}, L., {et~al.} 2017, \mnras, 470, 4231,
  \dodoi{10.1093/mnras/stx1440}

\bibitem[{{Safarzadeh} \& {Scannapieco}(2017)}]{saf}
{Safarzadeh}, M., \& {Scannapieco}, E. 2017, \apj, 850, 99,
  \dodoi{10.3847/1538-4357/aa94c8}

\bibitem[{{Saifollahi} {et~al.}(2021){Saifollahi}, {Trujillo}, {Beasley},
  {Peletier}, \& {Knapen}}]{Saifollahi+2021}
{Saifollahi}, T., {Trujillo}, I., {Beasley}, M.~A., {Peletier}, R.~F., \&
  {Knapen}, J.~H. 2021, \mnras, 502, 5921, \dodoi{10.1093/mnras/staa3016}

\bibitem[{{Sales} {et~al.}(2020){Sales}, {Navarro}, {Pe{\~n}afiel}, {Peng},
  {Lim}, \& {Hernquist}}]{Sales+2020}
{Sales}, L.~V., {Navarro}, J.~F., {Pe{\~n}afiel}, L., {et~al.} 2020, \mnras,
  494, 1848, \dodoi{10.1093/mnras/staa854}

\bibitem[{{Sawala} {et~al.}(2013){Sawala}, {Frenk}, {Crain}, {Jenkins},
  {Schaye}, {Theuns}, \& {Zavala}}]{sawala}
{Sawala}, T., {Frenk}, C.~S., {Crain}, R.~A., {et~al.} 2013, \mnras, 431, 1366,
  \dodoi{10.1093/mnras/stt259}

\bibitem[{{Schlegel} {et~al.}(1998){Schlegel}, {Finkbeiner}, \& {Davis}}]{SFD}
{Schlegel}, D.~J., {Finkbeiner}, D.~P., \& {Davis}, M. 1998, \apj, 500, 525,
  \dodoi{10.1086/305772}

\bibitem[{{Sif{\'o}n} {et~al.}(2018){Sif{\'o}n}, {van der Burg}, {Hoekstra},
  {Muzzin}, \& {Herbonnet}}]{sifon}
{Sif{\'o}n}, C., {van der Burg}, R. F.~J., {Hoekstra}, H., {Muzzin}, A., \&
  {Herbonnet}, R. 2018, \mnras, 473, 3747, \dodoi{10.1093/mnras/stx2648}

\bibitem[{{Spitler} \& {Forbes}(2009)}]{spitler}
{Spitler}, L.~R., \& {Forbes}, D.~A. 2009, \mnras, 392, L1,
  \dodoi{10.1111/j.1745-3933.2008.00567.x}

\bibitem[{{The Dark Energy Survey Collaboration}(2005)}]{des}
{The Dark Energy Survey Collaboration}. 2005, arXiv e-prints, astro.
\newblock \doarXiv{astro-ph/0510346}

\bibitem[{{Toloba} {et~al.}(2018){Toloba}, {Lim}, {Peng}, {Sales},
  {Guhathakurta}, {Mihos}, {C{\^o}t{\'e}}, {Boselli}, {Cuillandre},
  {Ferrarese}, {Gwyn}, {Lan{\c{c}}on}, {Mu{\~n}oz}, \& {Puzia}}]{toloba}
{Toloba}, E., {Lim}, S., {Peng}, E., {et~al.} 2018, \apjl, 856, L31,
  \dodoi{10.3847/2041-8213/aab603}

\bibitem[{{Tomozeiu} {et~al.}(2016){Tomozeiu}, {Mayer}, \& {Quinn}}]{tomozeiu}
{Tomozeiu}, M., {Mayer}, L., \& {Quinn}, T. 2016, \apj, 818, 193,
  \dodoi{10.3847/0004-637X/818/2/193}

\bibitem[{{Trujillo} {et~al.}(2020){Trujillo}, {Chamba}, \&
  {Knapen}}]{trujillo20}
{Trujillo}, I., {Chamba}, N., \& {Knapen}, J.~H. 2020, \mnras, 493, 87,
  \dodoi{10.1093/mnras/staa236}

\bibitem[{{van der Burg} {et~al.}(2016){van der Burg}, {Muzzin}, \&
  {Hoekstra}}]{vandenburg+2016}
{van der Burg}, R. F.~J., {Muzzin}, A., \& {Hoekstra}, H. 2016, \aap, 590, A20,
  \dodoi{10.1051/0004-6361/201628222}

\bibitem[{{van der Burg} {et~al.}(2017){van der Burg}, {Hoekstra}, {Muzzin},
  {Sif{\'o}n}, {Viola}, {Bremer}, {Brough}, {Driver}, {Erben}, {Heymans},
  {Hildebrandt}, {Holwerda}, {Klaes}, {Kuijken}, {McGee}, {Nakajima},
  {Napolitano}, {Norberg}, {Taylor}, \& {Valentijn}}]{vandenburg+2017}
{van der Burg}, R. F.~J., {Hoekstra}, H., {Muzzin}, A., {et~al.} 2017, \aap,
  607, A79, \dodoi{10.1051/0004-6361/201731335}

\bibitem[{{van der Walt} {et~al.}(2011){van der Walt}, {Colbert}, \&
  {Varoquaux}}]{numpy}
{van der Walt}, S., {Colbert}, S.~C., \& {Varoquaux}, G. 2011, Computing in
  Science and Engineering, 13, 22, \dodoi{10.1109/MCSE.2011.37}

\bibitem[{{van Dokkum} {et~al.}(2019{\natexlab{a}}){van Dokkum}, {Danieli},
  {Abraham}, {Conroy}, \& {Romanowsky}}]{vdkdm2}
{van Dokkum}, P., {Danieli}, S., {Abraham}, R., {Conroy}, C., \& {Romanowsky},
  A.~J. 2019{\natexlab{a}}, \apjl, 874, L5, \dodoi{10.3847/2041-8213/ab0d92}

\bibitem[{{van Dokkum} {et~al.}(2016){van Dokkum}, {Abraham}, {Brodie},
  {Conroy}, {Danieli}, {Merritt}, {Mowla}, {Romanowsky}, \&
  {Zhang}}]{vanDokkum+2016}
{van Dokkum}, P., {Abraham}, R., {Brodie}, J., {et~al.} 2016, \apjl, 828, L6,
  \dodoi{10.3847/2041-8205/828/1/L6}

\bibitem[{{van Dokkum} {et~al.}(2017){van Dokkum}, {Abraham}, {Romanowsky},
  {Brodie}, {Conroy}, {Danieli}, {Lokhorst}, {Merritt}, {Mowla}, \&
  {Zhang}}]{vdk17}
{van Dokkum}, P., {Abraham}, R., {Romanowsky}, A.~J., {et~al.} 2017, \apjl,
  844, L11, \dodoi{10.3847/2041-8213/aa7ca2}

\bibitem[{{van Dokkum} {et~al.}(2018{\natexlab{a}}){van Dokkum}, {Danieli},
  {Cohen}, {Merritt}, {Romanowsky}, {Abraham}, {Brodie}, {Conroy}, {Lokhorst},
  {Mowla}, {O'Sullivan}, \& {Zhang}}]{vdkdm1}
{van Dokkum}, P., {Danieli}, S., {Cohen}, Y., {et~al.} 2018{\natexlab{a}},
  \nat, 555, 629, \dodoi{10.1038/nature25767}

\bibitem[{{van Dokkum} {et~al.}(2018{\natexlab{b}}){van Dokkum}, {Cohen},
  {Danieli}, {Kruijssen}, {Romanowsky}, {Merritt}, {Abraham}, {Brodie},
  {Conroy}, {Lokhorst}, {Mowla}, {O'Sullivan}, \& {Zhang}}]{vdk-gc}
{van Dokkum}, P., {Cohen}, Y., {Danieli}, S., {et~al.} 2018{\natexlab{b}},
  \apjl, 856, L30, \dodoi{10.3847/2041-8213/aab60b}

\bibitem[{{van Dokkum} {et~al.}(2019{\natexlab{b}}){van Dokkum}, {Wasserman},
  {Danieli}, {Abraham}, {Brodie}, {Conroy}, {Forbes}, {Martin}, {Matuszewski},
  {Romanowsky}, \& {Villaume}}]{vdk19}
{van Dokkum}, P., {Wasserman}, A., {Danieli}, S., {et~al.} 2019{\natexlab{b}},
  \apj, 880, 91, \dodoi{10.3847/1538-4357/ab2914}

\bibitem[{{van Dokkum} {et~al.}(2022){van Dokkum}, {Shen}, {Keim},
  {Trujillo-Gomez}, {Danieli}, {Dutta Chowdhury}, {Abraham}, {Conroy},
  {Kruijssen}, {Nagai}, \& {Romanowsky}}]{vdk-bullet}
{van Dokkum}, P., {Shen}, Z., {Keim}, M.~A., {et~al.} 2022, \nat, 605, 435,
  \dodoi{10.1038/s41586-022-04665-6}

\bibitem[{{van Dokkum} {et~al.}(2015){van Dokkum}, {Abraham}, {Merritt},
  {Zhang}, {Geha}, \& {Conroy}}]{vanDokkum+2015}
{van Dokkum}, P.~G., {Abraham}, R., {Merritt}, A., {et~al.} 2015, \apjl, 798,
  L45, \dodoi{10.1088/2041-8205/798/2/L45}

\bibitem[{{Williams} {et~al.}(2004){Williams}, {Olszewski}, {Lesser}, \&
  {Burge}}]{90Prime}
{Williams}, G.~G., {Olszewski}, E., {Lesser}, M.~P., \& {Burge}, J.~H. 2004, in
  Society of Photo-Optical Instrumentation Engineers (SPIE) Conference Series,
  Vol. 5492, Ground-based Instrumentation for Astronomy, ed. A.~F.~M.
  {Moorwood} \& M.~{Iye}, 787--798, \dodoi{10.1117/12.552189}

\bibitem[{{Wolf} {et~al.}(2010){Wolf}, {Martinez}, {Bullock}, {Kaplinghat},
  {Geha}, {Mu{\~n}oz}, {Simon}, \& {Avedo}}]{wolf}
{Wolf}, J., {Martinez}, G.~D., {Bullock}, J.~S., {et~al.} 2010, \mnras, 406,
  1220, \dodoi{10.1111/j.1365-2966.2010.16753.x}

\bibitem[{{Wright} {et~al.}(2021){Wright}, {Tremmel}, {Brooks}, {Munshi},
  {Nagai}, {Sharma}, \& {Quinn}}]{Wright+2021}
{Wright}, A.~C., {Tremmel}, M., {Brooks}, A.~M., {et~al.} 2021, \mnras, 502,
  5370, \dodoi{10.1093/mnras/stab081}

\bibitem[{{Yagi} {et~al.}(2016){Yagi}, {Koda}, {Komiyama}, \&
  {Yamanoi}}]{Yagi+2016}
{Yagi}, M., {Koda}, J., {Komiyama}, Y., \& {Yamanoi}, H. 2016, \apjs, 225, 11,
  \dodoi{10.3847/0067-0049/225/1/11}

\bibitem[{{Zaritsky}(2022)}]{zgc}
{Zaritsky}, D. 2022, \mnras, 513, 2609, \dodoi{10.1093/mnras/stac1072}

\bibitem[{{Zaritsky} \& {Behroozi}(2023)}]{zb}
{Zaritsky}, D., \& {Behroozi}, P. 2023, \mnras, 519, 871,
  \dodoi{10.1093/mnras/stac3610}

\bibitem[{{Zaritsky} {et~al.}(2006){Zaritsky}, {Gonzalez}, \&
  {Zabludoff}}]{zfm1}
{Zaritsky}, D., {Gonzalez}, A.~H., \& {Zabludoff}, A.~I. 2006, \apj, 638, 725,
  \dodoi{10.1086/498672}

\bibitem[{{Zaritsky} {et~al.}(2008){Zaritsky}, {Zabludoff}, \&
  {Gonzalez}}]{zfm2}
{Zaritsky}, D., {Zabludoff}, A.~I., \& {Gonzalez}, A.~H. 2008, \apj, 682, 68,
  \dodoi{10.1086/529577}

\bibitem[{{Zaritsky} {et~al.}(2019){Zaritsky}, {Donnerstein}, {Dey},
  {Kadowaki}, {Zhang}, {Karunakaran}, {Mart{\'\i}nez-Delgado}, {Rahman}, \&
  {Spekkens}}]{Zaritsky+2019}
{Zaritsky}, D., {Donnerstein}, R., {Dey}, A., {et~al.} 2019, \apjs, 240, 1,
  \dodoi{10.3847/1538-4365/aaefe9}

\bibitem[{{Zaritsky} {et~al.}(2021){Zaritsky}, {Donnerstein}, {Dey},
  {Kadowaki}, {Zhang}, {Karunakaran}, {Mart{\'\i}nez-Delgado}, {Rahman}, \&
  {Spekkens}}]{Zaritsky+2021}
---. 2021, \apjs

\bibitem[{{Zaritsky} {et~al.}(2022){Zaritsky}, {Donnerstein}, {Dey},
  {Kadowaki}, {Zhang}, {Karunakaran}, {Mart{\'\i}nez-Delgado}, {Rahman}, \&
  {Spekkens}}]{Zaritsky+2022}
---. 2022, \apjs

\bibitem[{{Zou} {et~al.}(2017){Zou}, {Zhou}, {Fan}, {Zhang}, {Zhou}, {Nie},
  {Peng}, {McGreer}, {Jiang}, {Dey}, {Fan}, {He}, {Jiang}, {Lang}, {Lesser},
  {Ma}, {Mao}, {Schlegel}, \& {Wang}}]{Zou}
{Zou}, H., {Zhou}, X., {Fan}, X., {et~al.} 2017, \pasp, 129, 064101,
  \dodoi{10.1088/1538-3873/aa65ba}

\end{thebibliography}
\bibliographystyle{aasjournal}

\end{document}